\documentclass[prb, twocolumn, amsmath, amssymb]{revtex4}
\usepackage{graphicx}
\usepackage{bm}
\usepackage{color}

\begin{document}
\title{Ultrafast dynamics in relativistic Mott insulators}
\author{Jiajun Li}
\affiliation{University of Erlangen-Nuremberg, 91054 Erlangen, Germany}
\author{Nagamalleswararao Dasari}
\affiliation{University of Erlangen-Nuremberg, 91054 Erlangen, Germany}
\author{Martin Eckstein}
\affiliation{University of Erlangen-Nuremberg, 91054 Erlangen, Germany}

\begin{abstract}
We study the photoinduced ultrafast dynamics in relativistic Mott insulators, i.e., Mott insulators with strong spin-orbit coupling. For this purpose, we consider a minimal one-band Hubbard model on lattices with square and triangular symmetries, as relevant for layered transition metal compounds such as Sr$_2$IrO$_4$. Depending on the lattice and the spin-orbit coupling, the systems have canted antiferromagnetic or $120^\circ$ order. They are excited by simulating a short laser pulse, and the dynamics is solved using nonequilibrium dynamical mean-field theory. The pulse generates hot carriers, which subsequently perturb the magnetic order due to the coupling between the collective order and photocarriers. We find that this dynamics, which is known form regular antiferromagnets, depends sensitively on the spatial structure of the spin-orbit coupling. On the triangular lattice, in particular, relaxation times are influenced by the spin-orbit coupling for the chiral $120^\circ$ order, while on the square lattice with canted antiferromagnetic order the spin-orbit induced canting angle remains unchanged after the excitation. Our study opens up new possibilities of controlling magnetism and exotic spin states on the ultrafast timescales.
\end{abstract}

\maketitle

\section{Introduction}
%The ultrafast manipulation of collective orders in solids; ultrafast magnetism; theoretical understanding and significant promise in next-generation technologies.
%Control of material properties with nonequlibrium protocols emerged as an intriguing possibility in modern condensed matter physics \cite{basov2017}. 
The control of material properties with nonequlibrium protocols is one of the most fascinating promises of modern condensed matter physics \cite{basov2017,orenstein2012}. 
%\ME{Would it make sense to cite the Orenstein Reference here?}
%
One example is the ultrafast manipulation of 
%ME 
%the 
magnetic order, which can potentially boost major technological advances and reveal indispensable information of the underlying many-body physics \cite{kirilyuk2010,orenstein2012}. It has been known for decades that an ultrafast laser pulse can induce a subpicosecond demagnetization \cite{beaurepaire1996}, while a systematic theoretical description of the scenario is still under development. In many experimental studies, 
%ME
multi-temperature models are
%the multi-temperature model is 
adopted for the phenomenology, where the laser excitation suddenly injects energy into the electron reservoir in the metal, subsequently heating up and melting the collective magnetic order \cite{kirilyuk2010}. In insulating materials, this mechanism should nevertheless be modified \cite{kimel2002}, since the localized electrons cannot absorb energy effectively from the laser field. One possible mechanism is the delocalization of electrons due to the photoexcitation, termed \emph{photodoping}. The instantly created photocarriers then transfer their excess energy to the collective order and lead to the ultrafast melting. 

Of particular interest are Mott insulators, which are predicted to be metals from band theory but are indeed insulators due to the strong Coulomb repulsion \cite{imada1998}. The family 
%ME
includes 
%include 
various transition metal oxides, such as NiO, V$_2$O$_3$, and cuprates in certain parameter regimes \cite{Zhang2019,Held2001,Cai2016}. 
%ME
A minimal description is given by the one-band Hubbard model, 
%The prototypical example can be described as a one-band Hubbard model, 
%ME
with one orbital per each lattice site and one electron per orbital.
%where  there is one electron occupying one orbital per each lattice site. 
The strong onsite Coulomb repulsion $U$ prevents a double occupancy of the same site, leaving no room for electronic conduction. In these materials, the localized spins can order at low temperatures and tend to form an antiferrogmagnetic (AFM) phase 
%ME
on 
%in 
a bipartite crystal lattice, where neighboring spins align oppositely. The mechanism for the photo-induced melting of AFM order in Mott insulators is fundamentally different from the ultrafast demagnetization of ferromagnets.
%Photodoping of Mott insulators clearly reveals new scenarios for the ultrafast demagnetization, and has been experimentally studied in a variety of materials. 
It 
%ME
has been
%is 
shown that a laser pulse can quickly redistribute the charge among neighboring lattice sites, leading to fast generation of doubly occupied and empty sites (doublons and holons). 
%ME
These photocarriers can modify the electronic structure and the energy gap, 
%The energy gap and other parameters can be modified by the photocarriers, 
%ME
as observed, e.g., using 
%which have been observed using, e.g., 
angle-resolved photoemission spectroscopy (ARPES) and X-ray spectroscopy \cite{mor2017, beaud2014}.
%ME
The coupling between charge excitations and the spin order
%In particular, the coupling between charge excitations and the spin ordering 
has been argued 
%ME
to be
%as 
the dominant mechanism leading to the ultrafast spin dynamics \cite{lenarcic2013,Golez2014,balzer2015,DalConte2015}. 
%ME
Specifically, the motion of charge excitations can create trails of defects in the antiferromagnetic background, transferring energy from the hot photocarriers to the AFM order. The scenarios have been examined theoretically through a nonequilibrium generalization \cite{aoki2014} of the dynamical mean-field theory (DMFT) \cite{georges1996}, which is an established approach to describe the Mott physics in equilibrium. The same spin charge coupling becomes already manifest in the properties of a single hole or electron in the ordered spin background, which transforms into a spin-polaron, i.e., a hybrid of charge and magnon excitations\cite{Brinkman1970,Dagotto1994,Brunner2000,sangiovanni2006,Grusdt2018}. 
%Specifically, the motion of charge excitations can create trails of defects in the antiferromagnetic background \cite{Brinkman1970,Dagotto1994,Brunner2000,sangiovanni2006,DalConte2015,Grusdt2018}, transferring energy from the hot photocarriers to the AFM order and forming hybrid charge and magnon excitations, namely the spin-polarons. 
%ME

In equilibrium, it is known that the spin-polaron physics is strongly affected by geometric frustration \cite{Srivastava2005, Tohyama2006, Hamad2006, Hamad2008, Laeuchli2004, Shibata1999}, such as for the $120^\circ$ order in a triangular lattice \cite{Capriotti1999,White2007, Iida2019}. It should therefore be interesting to see how the mechanism for the melting of AFM order is modified in the presence of geometric frustration \cite{Bittner2020} or spin-orbit coupling (SOC). The latter can also be of interest for a practical reason: spin-orbit coupling can add a ferromagnetic moment to the AFM order, whose dynamics can be probed optically, while a  probe of pure antiferromagnetic correlations on the femtosecond timescale is more challenging \cite{afanasiev2019}.
%The spin-polaron physics can be enriched by spin frustration, such as for the $120^\circ$ order in a triangular lattice \cite{Capriotti1999,White2007, Iida2019}, leading to intriguing charge and order interplay \cite{Srivastava2005, Tohyama2006, Hamad2006, Hamad2008, Laeuchli2004, Shibata1999, Bittner2020}. The experimental verification of this demagnetization mechanism has, nevertheless, been challenging due to a lack of effective probe for the time evolution of antiferromagnetic correlation, which usually supports no macroscopic magnetic moment.
Recently, heavy transition metal compounds,
%ME
including elements such as iridium and ruthenium, 
%, e.g., with elements iridium and ruthenium, 
have attracted many research interests. 
%ME
Due to the large atomic number, the relativistic effect in these materials is significantly enhanced, and the spin-orbit coupling becomes of similar order of magnitude than other local electronic energy scales (Hund's coupling, crystal-field splitting, Coulomb interaction), leading to intriguing scenarios.
%The relativistic effect in these materials is significantly enhanced due to the large atomic number, leading to intriguing scenarios with strong spin-orbit coupling for the relevant electronic degrees of freedom. 
For example, Na$_2$IrO$_3$ and $\alpha$-RuCl$_3$ 
%ME
have been
%are 
predicted to realize the Kitaev honeycomb model in the low-energy sector,
%ME [???]
and may 
%and 
contain Majorana excitations promising for topological quantum computation \cite{jackeli2009}. The layered perovskite Sr$_2$IrO$_4$, on other hand, can be well described as a single-band Mott insulator \cite{jin2009,wang2011,kim2012}, but features an anomalously large ferromagnetic (FM) moment and potentially hosts states related to high-temperature superconductivity \cite{yan2015}. The photoinduced ultrafast dynamics in these materials has
%\ME{ I can't get rid of the feeling that ``the dynamics'' is singular, but I know it can be used both ways ...} 
been under intense scrutiny. With strong SOC, the nonthermal states created with photodoping can exhibit exotic properties that are absent for normal Mott insulators \cite{dean2016, versteeg2020}. In particular, the FM moment in Sr$_2$IrO$_4$ is rigidly coupled to the AFM order, which provides a unique opportunity for studying the ultrafast spin dynamics and the photodoping process by measuring the evolution of the FM moment \cite{afanasiev2019}. 
%ME
Microscopic 
%Although the experimental findings are promising, microscopic 
theories to describe these observations are scarcely available, partly due to the complex orbital degeneracy and the electron-lattice coupling, which place many challenges for theoretical studies. On the other hand, the strong SOC can split the energy band and sometimes gives rise to a single energy band near the Fermi energy, which can be effectively described by a one-band Hubbard model, as in the case of Sr$_2$IrO$_4$. This raises the question of how to understand the photoinduced spin dynamics in a minimal one-band spin-orbit Mott insulators.

In this article, we will concentrate on minimal one-band Hubbard models with SOC on a lattice of square or triangular symmetry, and use nonequilibrium DMFT to solve for the photoinduced dynamics. Two questions are mainly addressed here: 
%ME
(i) How can 
%(i) how could 
the presence of SOC affect the photoinduced spin dynamics in Mott insulators? (ii) Is there a general mechanism underpinning the dynamics in the different geometries? We will confirm that the (partial) melting of the spin order parameter is significantly affected by the strength of SOC, giving rise to a different pathway of tuning the ultrafast magnetic dynamics in solids and cold atom systems. 
%ME I would reverse the two sentences
The dynamics can be classified into two types, according to the spatial profile of the Dzyaloshinskii-Moriya interaction induced by the SOC. The effect of SOC on the ultrafast spin dynamics can be well understood by the spin polaron physics, i.e., the creation of magnons due to motion of charge excitations, in analogy to the case of normal Mott insulators without SOC.
%The effect of SOC on the ultrafast spin dynamics can be well understood by the spin polaron physics, i.e., the creation of magnons due to motion of charge excitations, in parallel to the magnetic melting in normal Mott insulators without SOC. The SOC-engineered spin dynamics can be further classified into two types, according to the spatial profile of the Dzyaloshinskii-Moriya interaction induced by the SOC. 

The article is organized as follows. Sect.~\ref{model} introduces the effective one-band Hubbard model with spin-orbit coupling and its solution within nonequilibrium DMFT. %Dasari nonequilibrium Dynamical Mean-Field Theory (DMFT). 
Sec.~\ref{mech} discusses the spin order under SOC and the different types of SOC Mott insulators. Sect.~\ref{sq} demonstrates the photoinduced spin dynamics in the prototypical spin-orbit Mott insulator Sr$_2$IrO$_4$ and discusses the strong coupling between the AFM order and the canting-induced FM moment. Sec.~\ref{tri} considers a triangular lattice with a different SOC pattern from the Sr$_2$IrO$_4$ case and shows the photoinduced demagnetization is qualitatively different. Sect.~\ref{conc} summarizes the main results and provides an outlook. %Sect.~\ref{conc} summarizes the article and provides an outllok.
%ME
%an
%
%outlook.

\section{Model and Method}
\label{model}
In 
%ME
%the 
multiband Mott insulators, the local spin orbit coupling $\bm{L} \cdot \bm S$ results in a rotation of the local spin-orbital basis. Up to the leading order, the intersite electron tunneling can mix up different spin and orbital indices when the local electronic Hamiltonian is diagonalized. As a minimal model, we consider the one-band spin-orbit Hubbard model 
%ME
with a spin-mixing hopping matrix,
%which includes a spin-mixing hopping matrix:
\begin{align}
H=-t_0\sum_{\langle ij\rangle\sigma\sigma'}e^{{\rm i}\varphi_{ij}(t)}R^{\langle ij\rangle}_{\sigma\sigma'}c^\dag_{i\sigma}c_{j\sigma'}+U\sum_i n_{i\uparrow} n_{i\downarrow},
\label{hgen}
\end{align}
%where $c$ represents the annihilation operator of the lattice electron and $U$ is the local 
%interaction parameter.
 where $c^{\phantom\dagger}_{i\sigma}$ represents the annihilation operator for an electron of spin $\sigma$ at the lattice site $i$ and $U$ is the on-site Coulomb interaction. The parameter $t_0$ characterizes the overall strength of electron tunneling between neighboring sites $\langle ij\rangle$, while the hopping matrix $\hat{R}^{\langle ij \rangle}$ determines how spin components are mixed during the tunneling process. $\varphi_{ij}(t) = e\bm A(t)\cdot (\bm r_i -\bm r_j)/\hbar$, with positions $\bm r_{i(j)}$ for two neighboring atoms, is the Peierls phase. It takes into account the impact of a laser pulse described by the time-dependent vector potential $\bm A(t)$, as in Fig.~\ref{soc_diag}(a). 
 
The simple model \eqref{hgen} can be implemented with atoms placed in optical lattices \cite{hamner2015}. It 
%ME
has also been
%is also 
argued that the two-dimensional spin-orbit Mott insulator Sr$_2$IrO$_4$ is described by a 
%ME
variant of 
%version of 
\eqref{hgen}. In this case, the strong spin-orbit coupling splits the $t_{2g}$ manifold of the $5d$ atomic shell, which are occupied by 5 electrons, leaving out a single-particle band with effective spin $J_{\rm eff}=1/2$ at the Fermi level \cite{jin2009}. In general, one can assume that the matrix $\hat{R}$ is proportional to a unitary rotation of the local spin-orbital basis.
%ME
In the remainder of this article, 
%In the rest of the article, 
we mainly consider hopping matrices in the following simple form,   
\begin{align}
\hat{R}^{ij}=\begin{pmatrix}e^{{\rm i}\phi_{ij}} & 0 \\ 0 & e^{-{\rm i}\phi_{ij}}\end{pmatrix}.
\label{rmatrix}
\end{align}
With time-reversal symmetry, this is the only possible form if one assumes 
%ME
that the 
$\hat{R}^{ij}$'s along different bonds can be diagonalized simultaneously. This assumption will not affect the general mechanism which will be discussed later. The spin-dependent hopping matrix \eqref{rmatrix} has a clear physical meaning: it measures the relative rotation between the local basis at site $i$ and $j$. Specifically, consider an electron at site $i$ with spin-state (a spinor) $\psi=\begin{pmatrix} c_\uparrow,c_\downarrow\end{pmatrix}^T$, the hopping process to site $j$ transforms the state into $\hat{R}^{ij}\psi=\begin{pmatrix} e^{{\rm i}\phi_{ij}}c_\uparrow, e^{-{\rm i}\phi_{ij}}c_\downarrow\end{pmatrix}^T$.

The SOC angle $\phi_{ij}$ generally depends on the bond $\langle ij\rangle$. In different lattices and with different patterns of $\phi_{ij}$, the physics can differ dramatically, see Fig.~\ref{soc_diag}(b-d) for examples. Here we exemplarily consider a square lattice and a triangular lattice. These lattices are generally related to various transition metal compounds, such as Sr$_2$IrO$_4$, the copper-oxide monolayer \cite{bonesteel1992} and Na$_2$IrO$_3$ \cite{shitade2009}. When one electron hops around a loop in the lattice, it captures an overall phase factor $\sigma\sum\phi_{ij}$ accumulating the SOC phases through the trajectory. The phase differs 
%ME
by 
a sign $\sigma=\pm1$ for effective spin up and down, respectively. As shown in Fig.~\ref{soc_diag}, the square lattice can have two different SOC patterns, featuring zero or spin-dependent $4\sigma\phi$ flux for each unit cell. Such different patterns imply different magnetic orders, which is best understood in terms of the strong-coupling spin model obtained from the Hubbard model (see Sec.~\ref{mech}). Indeed, the single-band effective model of Sr$_2$IrO$_4$ \cite{jackeli2009} features the zero flux pattern.
%ME
This leads to a canted antiferromagnetic (AFM) phase with a ferromagnetic (FM) moment. 
%, which leads to a canted antiferromagnetic (AFM) phase with a ferromagnetic (FM) moment. 
Cuprates, on the other hand, are believed to have the $4\sigma\phi$ SOC flux and frustrated spin canting \cite{bonesteel1992}. Another example of a model with nonzero SOC flux is the triangular lattice shown in Fig.~\ref{soc_diag}(d), which will be discussed in more detail in the following sections. We shall see that the two scenarios with zero or nonzero SOC flux lead to distinct photoinduced spin dynamics.

\subsection{DMFT solution of the dynamics}
We will study the photoinduced spin dynamics of model~\eqref{hgen} using 
nonequilibrium DMFT \cite{aoki2014}. Specifically, a short electric pulse is applied to the system, described by a vector potential of the following form, 
\begin{align}
\bm{A}(t) = A_0 e^{-(t-\tau_0)^2/2\Sigma^2}\sin\left(\frac{2\pi}{T} t\right)\bm e_p,
\label{pulse}
\end{align}
where $\bm e_p$ is a unit linear-polarization vector and is chosen to be along the diagonal, which means $(\sqrt{2}/2,\sqrt{2}/2)$ for the square lattice and $(\sqrt{3}/2,1/2)$ for the triangular lattice. In DMFT, the lattice model is mapped to a single-impurity Anderson model where a lattice site is coupled to a Fermion reservoir. The hybridization function of the reservoir $\Delta_{\sigma\sigma'}(t,t')$ is self-consistently determined by the lattice environment connected to the impurity site. The auxiliary impurity model is solved with a strong coupling expansion up to the lowest order (the non-crossing approximation) \cite{schuler2020}.

To reduce the computational costs and access the dynamics for longer time, we study the model on infinitely connected Bethe lattices in which the local environment of each lattice site equals the environment of a site on the square or triangular lattice.
%ME
Quite generally, on the infinitely connected Bethe lattice the hybridization-function on a site $j$ is given by $\hat{\Delta}_j(t,t')= \sum_{j'} T_{j,j'}(t) \hat{G}_{j'}(t,t')T_{j',j}(t')$, where the sum runs over all neighbors $j'$ of $j$, $T_{j,j'}$ is the tunnelling matrix element between neighbouring  sites $j$ and $j'$, and $\hat{G}_{j'}(t,t')$ is the local Green's function on site $j'$. This can be readily generalized to the case in which the hopping is a matrix in spin space and carries a Peierls phase, and $\hat{G}_{j'}(t,t')$ depends on the sublattice on which $j'$ is located. 

The simple square lattice is bipartite, so we generally solve two impurity problems with hybridization functions 
%ME
$\hat{\Delta}^{1,2}(t,t')=\frac{t_0^2}{4} \sum_\alpha e^{-{\rm i}\varphi_\alpha(t)}\hat{R}^{\alpha\dag}\hat{G}^{2,1}(t,t')\hat{R}^{\alpha}e^{{\rm i}\varphi_\alpha(t')}$. 
%$\hat{\Delta}^{1,2}(t,t')=\frac{t_0^2}{4} \sum_\alpha e^{i(\varphi_\alpha(t)-\varphi_\alpha(t'))}\hat{R}^{\alpha\dag}\hat{G}^{2,1}(t,t')\hat{R}^{\alpha}$. 
The two impurities $1,2$ correspond to two sublattices in the case of a staggered AFM order, and the $\hat{G}^{1,2}$ is the local Green's function for the impurity $1,2$, respectively. The index $\alpha$ labels 4 types of bonds connecting to each site in a Bethe lattice: the hopping matrices $R^\alpha$ and Peierls phases $\phi_\alpha(t)$ are chosen according to the four different bonds adjacent to each site in the square lattice. In particular, the Perierls phase $\phi_\alpha$ is obtained by projecting the vector potential given by Eq.~\eqref{pulse} on the four different directions of the square lattice.

\begin{figure}
\includegraphics[scale=2.2]{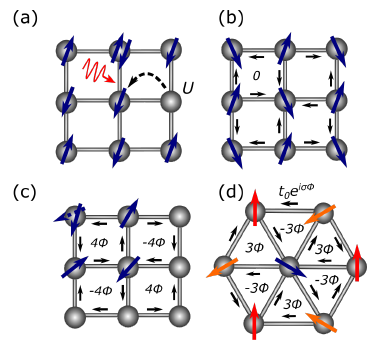}
\caption{Photoexcitation and 
%ME
%the 
spin-orbit coupling (SOC) patterns in the one-band Hubbard model. (a) Photoexcitation of a collinear antiferromagnetic ground state. Doublons and holons are instantly created by the laser excitation and can hop around the lattice, leaving out traces of defects in the AFM background. (b-d) Possible patterns of the 
%ME
spin-orbit angles 
$\phi_{ij}$ and 
%ME
the corresponding 
spin orders in two-dimensional lattices, including the square lattice with (b) SOC flux $0$ and (c) SOC flux $4\sigma\phi$ and (d) triangular lattice with flux $3\sigma\phi$. In (c), the dashed arrow indicates the spin canting is frustrated. In (d), the three spin directions (sublattices 1, 2, 3) for the $120^\circ$ AFM order are indicated with red, orange, and blue colors.  The black arrow indicates the direction in which a hopping gives rise to a phase factor $e^{{\rm i}\sigma\phi}$ for spin index $\sigma=\pm1$. Hopping in the reversed direction gives $e^{-{\rm i}\sigma\phi}$ correspondingly. 
%ME
%Red arrows indicate the possible spin orientations. 
}
\label{soc_diag}
\end{figure}

For the triangular 
%ME
case,
%lattice, 
a similar self-consistent equation can be written out as above. 
%ME
We consider three impurities for the three inequivalent sites on the lattice, 
as in the triangular lattice with $120^{\circ}$ order. A site of type $1$ has three neighbors
each on sublattice $2$ and $3$, as seen in Fig.~\ref{soc_diag}. The hybridization is given by $\hat{\Delta}^{1}(t,t')=\frac{t_0^2}{6} \sum_\alpha [e^{-{\rm i}(\varphi_\alpha(t)-\varphi_\alpha(t'))}\hat{R}^{\alpha\dag}\hat{G}^{2}(t,t')\hat{R}^{\alpha}+e^{{\rm i}(\varphi_\alpha(t)-\varphi_\alpha(t'))}\hat{R}^{\alpha}\hat{G}^{3}(t,t')\hat{R}^{\alpha\dag}]$, where $\alpha$ represents the three different bonds. Note that the phase of the $R$ matrices  for hopping from $1$ and $2$ are opposite to those for hopping from $1$ and $3$. The corresponding expression for the hybridization function on the impurities $2$ and $3$ is obtained by a cyclic permutation of the indices. 
%The main difference is that we consider three impurities for including the frustration effect. For impurity $1$, we impose the following condition $\hat{\Delta}^{1}(t,t')=\frac{t_0^2}{6} \sum_\alpha [e^{i(\varphi_\alpha(t)-\varphi_\alpha(t'))}\hat{R}^{\alpha\dag}\hat{G}^{2}(t,t')\hat{R}^{\alpha}+e^{-i(\varphi_\alpha(t)-\varphi_\alpha(t'))}\hat{R}^{\alpha}\hat{G}^{3}(t,t')\hat{R}^{\alpha\dag}]$. Here, $\alpha$ represents the three different bonds in a triangular lattice. The analogous conditions are imposed for impurity $2,3$. 
In the typical case of planar $120^\circ$ spin order as shown in Fig.~\ref{soc_diag}(d), the Green's functions $\hat{G}^s$ ($s=1,2,3$) are connected by subsequent $2\pi/3$ rotations: $\hat{G}^1=\hat{U}^\dag(\pm2\pi/3)\hat{G}^2\hat{U}(\pm2\pi/3)=\hat{U}^\dag(\pm4\pi/3)\hat{G}^3\hat{U}(\pm4\pi/3)$, where the $\pm$ sign leads to different chiralities of the order and is not important in the present study \cite{kawamura1998}. The operator $\hat{U}(\theta)=\exp(i\hbar\tau^z \theta/2)$ is a rotation around the spin $J^z$--axis, where $\tau^z$ is the usual Pauli matrix. This arrangement essentially represents a triangular loop of impurities.
 
\section{The spin order in relativistic Mott insulators}
\label{mech}
%We have seen two types of phenomenology for the magnetic dynamics with different SOC patterns. With the spin-orbit $\hat{R}$ matrix given by \eqref{rmatrix}, depending on whether it can be eliminated by a unitary transformation, the physical consequences are also distinct. In this section, we clarify the general principle underlying the classification. Indeed, the SOC patterns can be categorized according to whether the SOC-induced \emph{spin canting} is frustrated. Furthermore, we will see that with frustrated spin canting, the spin exchange energy is generally modified, leading to enhanced or suppressed spin polaron peaks and photoinduced demagnetization. 

%\subsection{The SOC-induced Dzyaloshinskii-Moriya interaction}
We first study the spin order in the spin-orbit Mott insulators. For later convenience, we define the spin operator $J^\mu=\sum_{\sigma\sigma'}c^\dag_\sigma\tau^\mu_{\sigma\sigma'}c_\sigma'$ in the single-band model. While $\bm J$ corresponds to a total angular momentum state and is often referred to as ``effective spin" in the literature, we will use the short-hand ``spin" for simplicity in the following. However, it is worth noting that the effective spin is generally related to the experimentally observed magnetic moments $\bm M$ through a tensor relation \cite{wang2011}. 

In the strong-coupling limit $U\gg t_0$, the low-energy physics of \eqref{hgen} is captured by a spin exchange model obtained from the Schrieffer-Wolff transformation, while the spin-mixing $\hat{R}$-matrix leads to a Dzyaloshinskii-Moriya (DM) interaction \cite{moriya1960,jackeli2009}. For example, consider $\phi=0$, a half-filled two-site Hubbard chain with site index $i=1,2$ is effectively described by the following Heisenberg model,
\begin{align}
I_{\rm ex}\bm J_{1}\cdot \bm J_{2},
\label{heisenberg}
\end{align}
where the exchange coupling $I_{\rm ex}=4t_0^2/U$ as usual. At low temperatures, this Hamiltonian gives rise to an antiferromagnetic ground state. In the presence of SOC, a nonzero phase $\phi$ generally leads to a Dzyaloshinskii-Moriya (DM) interaction. The easiest way to understand the situation is to impose a local basis rotation $c^\dag_{j\sigma}=e^{{\rm i}(-1)^j\sigma\phi/2}\tilde{c}^\dag_{j\sigma}$, with $\sigma=\pm1$ for up/down spin, mapping the model to the $\phi=0$ case. The mapping rotates the local spins in the following way
\begin{align}
&\tilde{J}_{j}^x=J_{j}^x\cos{\left[(-1)^j\phi\right]} - J_{j}^y\sin{\left[(-1)^j\phi\right]}, \nonumber\\
&\tilde{J}_{j}^y=J_{j}^x\sin{\left[(-1)^j\phi\right]} + J_{j}^y\cos{\left[(-1)^j\phi\right]},\nonumber\\
&\tilde{J}_{ j}^z=J_{j}^z,
\label{rot}
\end{align}
 and transforms the effective hamiltonian \eqref{heisenberg} to a model with a DM interaction,
\begin{align}
&I_{\rm ex}\tilde{\bm J}_{1}\cdot \tilde{ \bm J}_{2}=I_{\rm ex}\cos(2\phi) ( J^x_{1} J^x_{2}+J^y_{1} J^y_{2}) \nonumber\\
&\quad- I_{\rm ex} J^z_{1} J^z_{2}+I_{\rm ex}\sin(2\phi) \bm e_z\cdot (\bm J_{1}\times \bm J_{2}).
\label{dm}
\end{align}
Note that, for a two-site model, a general $\hat{R}$ matrix can always be recast to the form in~\eqref{rmatrix} through redefining the $J^z$--axis. 

The last term in Eq.~\eqref{dm}, or the DM interaction, tends to induce a relative rotation between two neighboring spin moments, which is always possible for the two-site model. However, in a lattice with a certain $\phi_{ij}$ pattern, the DM interaction from different bonds adjacent to a given spin can favor inconsistent canting angles, resulting in \emph{frustration} of the spin canting \cite{moriya1960, bonesteel1992}. The diagrams in Fig.~\ref{soc_diag}(b) and (c) show two examples of the unfrustrated (with flux $0$) and frustrated (with flux $4\sigma\phi$) spin canting for a square lattice. The frustration corresponds to 
%ME 
a
%the 
nonzero SOC flux for a closed loop in the lattice. Indeed, if the SOC flux %ME
vanishes for all loops in the lattice, 
%for any loops in the lattice identically vanish, 
one can 
%ME
%then 
always impose the above-mentioned local basis rotation to map the full SOC Hubbard model to a non-SOC Hubbard model with $\phi=0$, thus completely eliminating the DM interaction. In this case, the effect of SOC is simply inducing a spin canting, and the resulting spin order is equivalent to the original one up to the local basis rotation. In the following, we consider two examples: Sr$_2$IrO$_4$, which corresponds to a lattice of square symmetry with zero SOC flux, and a lattice of triangular symmetry with nonzero flux.

\subsection{The canted AFM order of Sr$_2$IrO$_4$}
The case of Sr$_2$IrO$_4$ is an example of unfrustrated spin canting on a square lattice. In the following, we will restrict the discussion to consider the AFM order inside the spin $xy$--plane, as found in \emph{ab initio} calculations \cite{jin2009}. Specifically, the SOC angle is given as follows: for bond $\langle ij\rangle$ one has $\phi_{ij}=\zeta_i\phi$, where $\zeta_i = \pm1$ for the two sublattices \cite{jin2009}. A unitary transformation $\hat{S}=e^{{\rm i}\phi\sum_i\sigma\zeta_i  c^\dag_{i\sigma}c_{i\sigma}/2}$ can rotate the phase of the local operators $c,c^\dag$ like $\tilde{c}_{i\sigma}=\hat{S}^\dag c_{i\sigma}\hat{S}=e^{{\rm i}\sigma\zeta_i\phi/2}c_{i\sigma}$, such that the hopping part of the Hamiltonian in Eq.~\eqref{hgen} is recast into $H_0=-t_0\sum_{\langle ij\rangle\sigma} e^{{\rm i}\varphi_{ij}(t)}\tilde{c}^\dag_{i\sigma}\tilde{c}_{j\sigma}$. Note that the transformation $\hat{S}$ is indeed a rotation of angle $\phi$ around the $J^z$--axis in the effective spin space (see Eq.~\eqref{rot}). The original Hamiltonian is mapped to a Hubbard model without SOC \cite{wang2011}. Therefore, one can conclude the spin order of Sr$_2$IrO$_4$ is a canted AFM with canting angle $\phi$, which is rotated to a collinear AFM of $\tilde{\bm J}$ under the above transformation. This is a concrete example of the general principle presented in the beginning of the section \cite{jackeli2009}. 

Note that the $\phi$--rotation is static and preserved by time evolution. During the photodoped dynamics considered in the following, the same transformation should also map the full dynamics of $J^\mu$ to the evolution of collinear $\tilde{J}^\mu$. Therefore, in the one-band model, the whole physics of SOC-induced spin canting is explained with a static rotation of the local spin basis of angle $\phi$. This immediately gives rise to two important consequences which we will confirm in the DMFT simulation: (i) If $t_0$ is fixed and $\phi$ is changed, the photodoped dynamics should be unaffected at all. (ii) During the photoinduced dynamics, the canting angle of the AFM order should always be $\phi$, implying a robust coupling between the AFM and the FM moments. 

\subsection{Chiral AFM order in a triangular lattice}
Since the triangular lattice is not bipartite, a collinear AFM order is impossible, while a $120^\circ$ canted antiferromagnetic order has been confirmed to exist by both theoretical and experimental studies \cite{Capriotti1999,White2007, Iida2019}. In this case, three atoms in the same triangle align their (effective) spins $120^\circ$ from each other, as shown in Fig.~\ref{soc_diag}(d).
The SOC pattern shown in the same figure leads to completely frustrated spin canting and cannot be eliminated by a unitary transformation. 

A crucial nature of the $120^\circ$ AFM order is its \emph{chirality} \cite{kawamura1998}. In particular, when one traverses the three vertices of a triangular plaquette in the counterclockwise manner, the subsequently encountered spins are rotated relative to each other either counterclockwise or clockwise, corresponding to two chiral AFM orders. For example, one can define the chirality with respect to the triangles labelled by flux $+3\phi$; the order in Fig.~\ref{soc_diag}(d) then represents the counterclockwise case, and the clockwise case can be obtained by reversing all three spins in the same diagram. The two chiral orders are equivalent without SOC. However, in the presence of SOC ($\phi\ne0$), the situation becomes different. Indeed, the SOC-induced DM interaction tends to stabilize a certain relative angle between two neighboring spins. Depending on whether this DM-favored angle is consistent with the chiral $120^\circ$ order, the total energy can be either lowered or elevated, respectively. Moreover, as we shall see later, the motion of charge excitations, i.e., doubly occupied sites and empty sites, in the AFM background plays a crucial role for the ultrafast dynamics. The SOC flux modifies the effective potential felt by charge excitations, affecting the time evolution of the spin order.

Finally we note that, although we have considered a special type of the SOC matrix $\hat{R}$ in \eqref{rmatrix}, the principle itself can be widely applicable. For example, when $\hat{R}^{\hat{x}}$ and $\hat{R}^{\hat{y}}$ do not commute, such as for Rashba-type SOC in two-dimensional lattices, the phenomenology should fall into the category of frustrated spin canting since for a closed loop $\hat{R}^{\hat{y}\dag}\hat{R}^{\hat{x}\dag}\hat{R}^{\hat{y}}\hat{R}^{\hat{x}}\ne \mathbb{I}$ in general.

\section{Ultrafast dynamics in {S\lowercase{r}$_2$I\lowercase{r}O$_4$}}
\label{sq}

%\ME{Should we highlight more from the outset  directly that this is really close to the Afanasiev paper, providing a more detailed analysis of the angle dependence? }
In this section, we consider the photoinduced melting of the spin order in Sr$_2$IrO$_4$ using the single-band model. As explained above, the effective model of Sr$_2$IrO$_4$ features an SOC pattern with zero SOC flux, leading to a ferromagnetic moment consistent with the DM interaction of bonds in different directions. In this case, the SOC phase factor can be completely gauged away with a unitary transformation, and the photoinduced dynamics should also be unaffected with or without the spin-orbit coupling. The ultrafast dynamics of Sr$_2$IrO$_4$ has been experimentally studied in a recent work \cite{afanasiev2019}, which contains a detailed analysis of the evolution of the canting angle. 

In a realistic description, the SOC angle $\phi$ is nevertheless controlled by the rotation angle $\theta$ of the IrO$_6$ octahedra \cite{jin2009}, which simultaneously affects the electron tunneling parameter $t_0$ in the following way, 
\begin{align}
t_0\hat{R}^{\hat{x}/\hat{y}}=\frac{2\bar{t}_0}{3}[\cos\theta (2\cos^4\theta-1)+i\sin\theta (2\sin^4\theta-1)\tau^z],
\label{t0}
\end{align}
where 
%ME
$\tau^z=\begin{pmatrix}1 & 0 \\ 0 & -1 \end{pmatrix}$,
%$\tau^z=\begin{pmatrix}1 & 0 \\ 0 & -1 \end{pmatrix}$ 
and $\theta=0$ corresponds to the non-SOC case, see Fig.~\ref{sq_canting}(a). In experiments, the angle $\theta$ is 
%ME
%apparently 
controllable through a strain \cite{liu2015}. 

In the following, we address the question of how the spin dynamics can be controlled by changing the structural parameter $\theta$. We choose $\bar{t}_0=1.0$ in Eq.~\eqref{t0}, which can be viewed as the energy unit, and the time unit is $\hbar/\bar{t}_0$ with $\hbar=1$. For the interaction parameter, we pick up $U=4.5\bar{t}_0$, which is close to the realistic parameters in Sr$_2$IrO$_4$. We vary $\theta=5^\circ, 10^\circ,\ldots, 30^\circ$ for demonstrating the effect of SOC \cite{afanasiev2019}. $\theta\sim 11^\circ$ is reported in experiments. The half-filling condition $n_\uparrow+n_\downarrow=1$ is imposed with a chemical potential $\mu=U/2$, yielding an insulating ground state with AFM ordering. % With the Bethe-lattice self-consistency, the bandwidth is $W=4t_0$. 
 An electric pulse, as given by Eq.~\eqref{pulse}, is applied at $t=0$ with parameters $\Sigma = 0.5, \tau_0=5.0, A_0=0.8, T=4.0$. Since we are only concerned with the spin melting following the excitation, the detail of the protocol should not matter too much.
\begin{figure}
\includegraphics[scale=0.7]{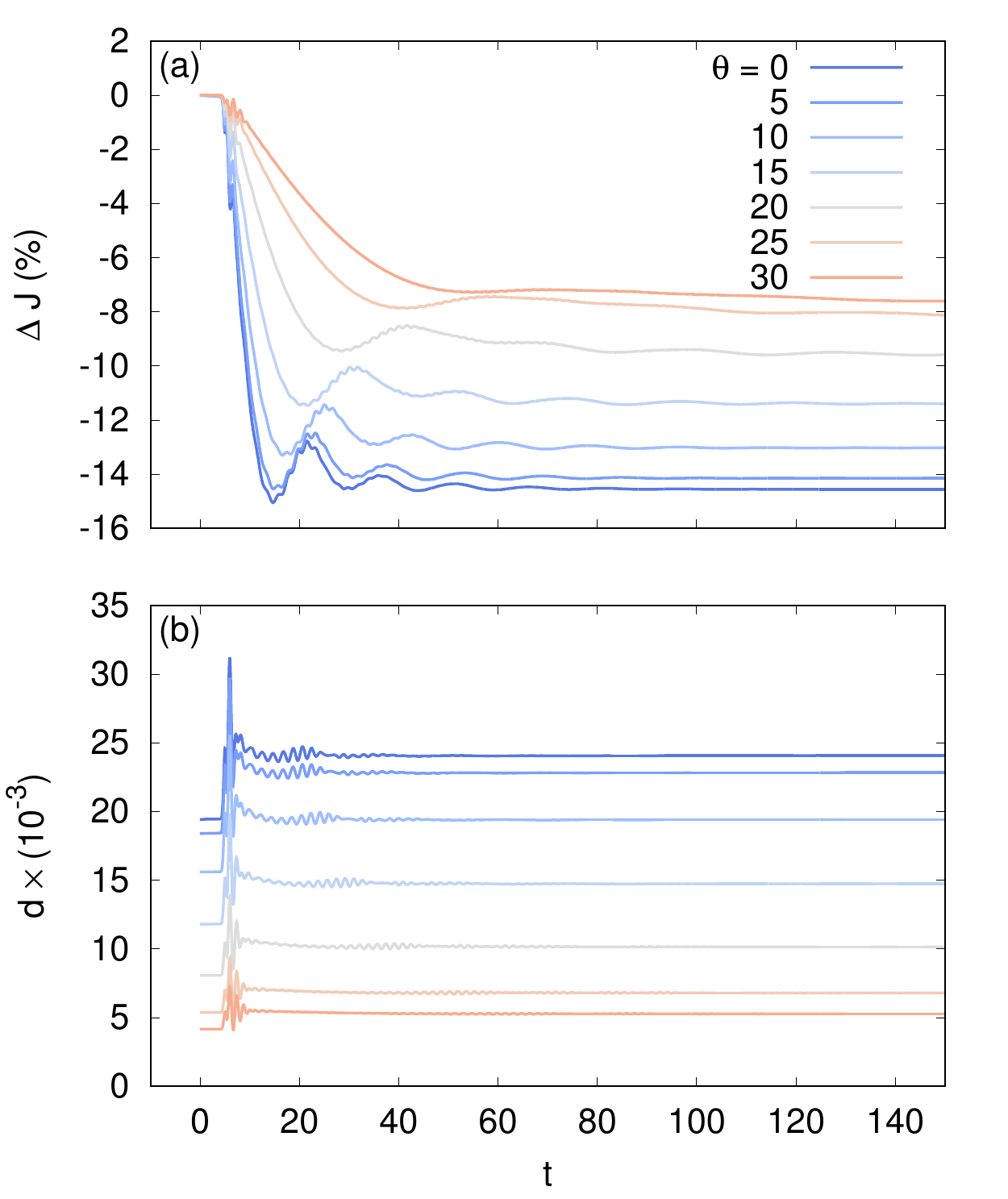}
\caption{The photoinduced dynamics in the one-band Hubbard model of Sr$_2$IrO$_4$ for different octahedra rotation $\theta$. (a) The melting of the spin order parameter $J$. (b) The time evolution of the double occupancy.  An initial inverse temperature $\beta=50$ is used.}
\label{sq_decay}
\end{figure}

\begin{figure}
\includegraphics[scale=0.7]{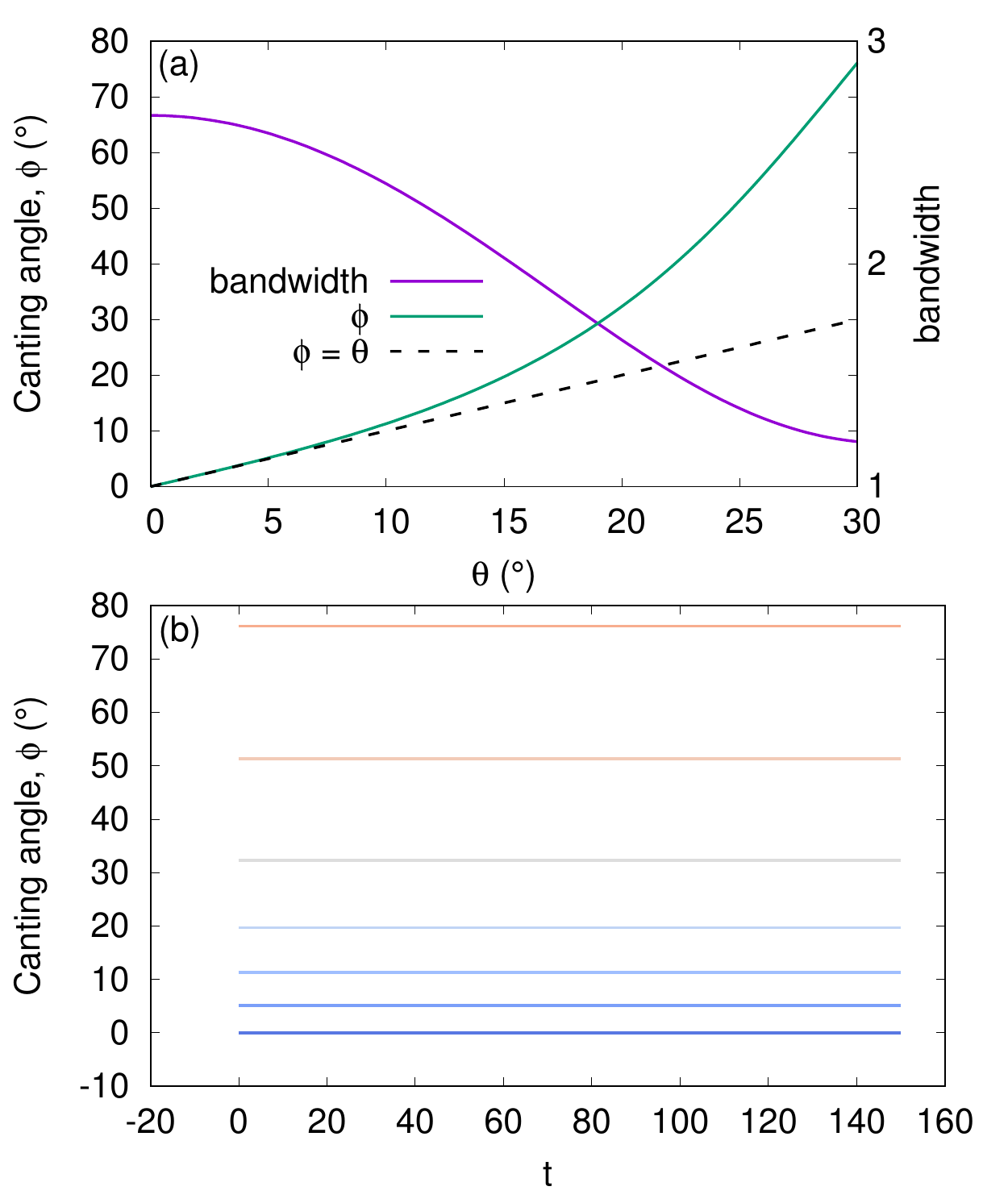}
\caption{(a) The dependence of the canting angle $\phi$ and non-interacting bandwidth $4t_0$ on the structural parameter $\theta$.  (b) The time evolution of the canting angle after the ultrafast electric pulse.  Parameters are identical to Fig.~\ref{sq_decay}.}
\label{sq_canting}
\end{figure}

\subsection{Demagnetization of the canted AFM order}

In equilibrium, the model yields a canted antiferromagnetic order with canting angle $\phi$. For small $\theta$ parameter, we have $\phi\sim\theta$. We suppose the AFM is measured by a staggered $J^y$ and the canting-induced ferromagnetic moment is $J^x$. Then the canting angle is given by $\phi=\arctan (J^x/J^y)$. The pulse-induced spin dynamics with different $\theta$'s are shown in Fig.~\ref{sq_decay}(a), indicating an ultrafast melting of the spin order parameter $|\bm J|$ and thus the ordered magnetic moment $|\bm M|$. At the same time, the photoexcitation creates charge excitations, i.e. doublons and holons (empty sites), in the Mott insulator, which slowly recombine on larger timescales. As shown in Fig.~\ref{sq_decay}(b), $d=\langle n_{\uparrow}n_{\downarrow}\rangle$, measuring the amount of doubly occupied sites (doublons, $\left|\uparrow\downarrow\right\rangle$), is transiently enhanced by the pulse and then decays. 

One important observation regarding the melting of the spin order is that it happens over a timescale much longer than the duration of the pulse. Indeed, the femtosecond pulse mainly injects energy into the charge excitations, and the subsequent melting of the collective order is  related to the coupling between the antiferromagnetic order and the photocarriers \cite{lenarcic2013}. When the charge excitations hop on the lattice, they continuously create traces of defects in the AFM background and destroy the order \cite{balzer2015}. This process is illustrated in Fig.~\ref{soc_diag}(a), where the motion of a holon creates several mismatched bonds in the AFM background. During the process, the photoexcited system evolves into a long-lived photodoped state, in which double occupancy evolves slowly, indicating a slow recombination of doublons and holons \cite{iwai2003, okamoto2010, sensarma2010, eckstein2014, mitrano2014}.

 %This is exactly what we expect from the analytical understanding in Sec.~\ref{mech}. 
A small angle $\theta\sim 5^\circ$ only minimally changes the dynamics, while a large $\theta\gtrsim 15^\circ$ strongly suppresses the partial melting (Fig.~\ref{sq_decay}(a)). The same trend is observed for $\theta <0$. As discussed in Sec.~\ref{mech}, if one would fix $t_0$ and only change $\phi$, the spin dynamics would be unaffected and always identical to the $\phi=0$ curve in Fig.~\ref{sq_decay}. The dependence of the dynamics on $\theta$ is explained by the modification of the bandwidth when $\theta$ is changed. It can be seen from Fig.~\ref{sq_canting}(a) that the non-interacting bandwidth $4t_0$ monotonically decreases with increasing $\theta$ up to about $\theta\approx 30^\circ$, although the nominal $\bar{t}_0$ is fixed. Larger $U/t_0$ generally results in a larger Mott gap, so the pulse creates fewer charge excitations in the system, leading to weaker melting of the spin order. During the spin dynamics, a prominent feature is that the canting angle remains fixed, as shown in Fig.~\ref{sq_canting}(b), implying a very robust coupling between the AFM and FM moments. This phenomenon has been explained by the discussion in Sec.~\ref{mech}. The fixed canting angle provides a unique opportunity to study the ultrafast spin dynamics because the FM moment is easier to measure in experiments \cite{afanasiev2019}.  This robust coupling between the AFM and FM moments may, nonetheless, be broken by the electron-phonon coupling which can dynamically modify the structural parameter $\theta$, and the multi-band physics which is not included in the present picture. 

%The numerical results discussed above suggest a promising pathway of controlling spin dynamics using SOC. It can be speculated that a carefully imposed strain can modify the structural distortion, such as the IrO$_6$ octahera rotation in the case of Sr$_2$IrO$_4$, tuning the bandwidth and/or the SOC angle $\phi$, and providing a control knob for the photoinduced spin dynamics in spin-orbit Mott insulators.
%The above numerical results suggest an experimental route to further confirm the rigid coupling of AFM and FM moments without directly measuring the dynamics of the AFM: By imposing a strain, one can change the canting angle. If for each canting angle the amplitude of the pump laser is adjusted to yield the same excitation density, the demagnetization time should be fixed. 

\section{Ultrafast dynamics in a frustrated Mott insulator}
\label{tri}

In this section, we turn to the spin-orbit Hubbard model with the triangular geometry. In this case, we set $t_0=1.0$ as an energy unit and directly vary the SOC angle $\phi$. With the DMFT calculations, the $120^\circ$ AFM has been found in the ground state at inverse temperature $\beta=50$. The spin canting is completely frustrated in this case, unless $\phi=120^{\circ}$, which amounts to a spatial reflection and is not important in the present study. We will be mostly interested in $\phi\ne120^\circ$, in particular smaller values of $\phi$. 

As mentioned above, both the $120^\circ$ AFM order and the SOC pattern can have two distinct chiralities. In the following discussion, we will fix the chirality of the order to be counterclockwise when traversing sublattices in the order 1 to 2 to 3 (see Fig.~\ref{soc_diag}(d)), and change $\phi$ for positive \emph{and} negative values.

\begin{figure}
\includegraphics[scale=0.7]{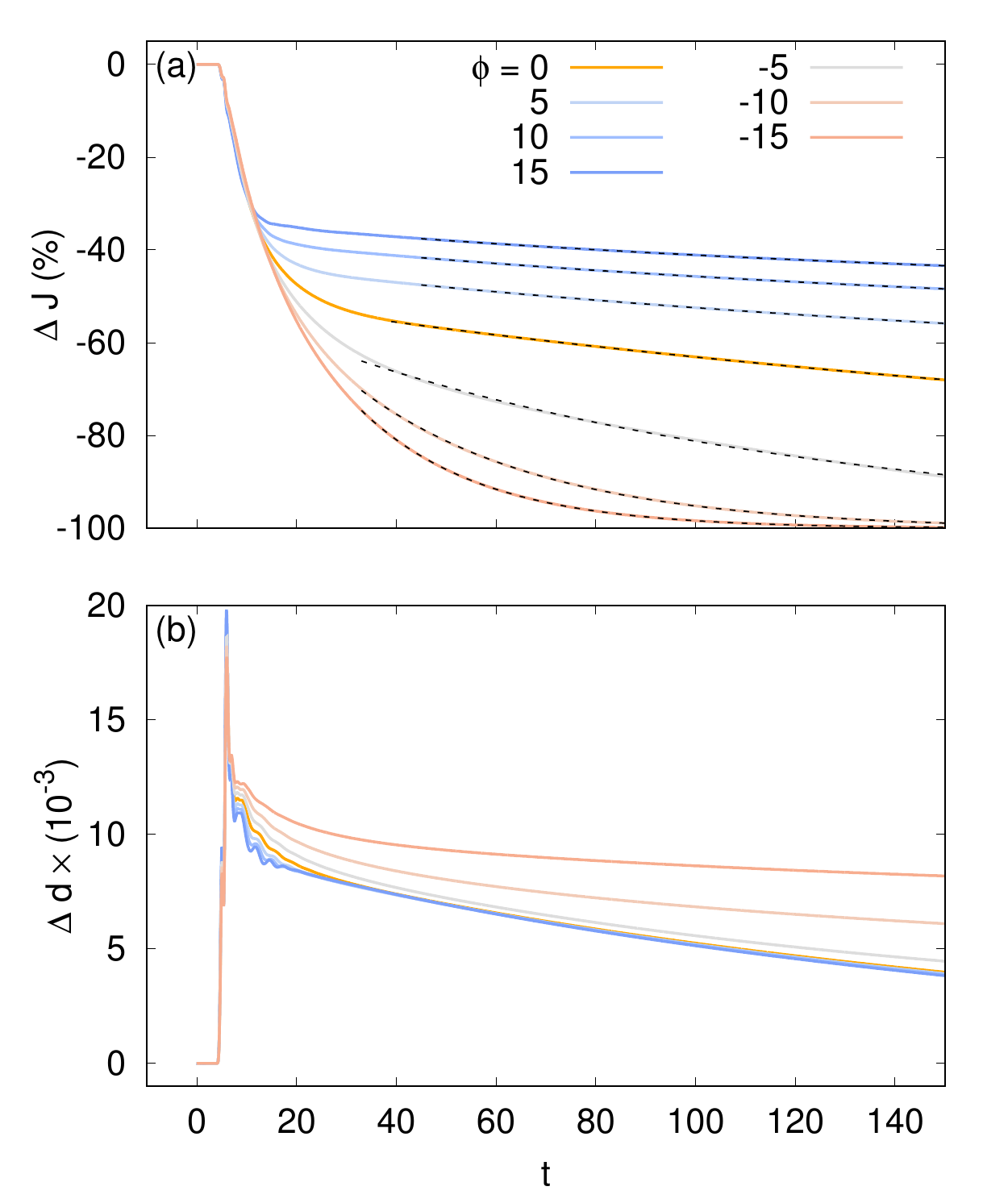}
\caption{The photoinduced dynamics in the frustrated SOC Hubbard model for different SOC parameter $\phi$. (a) The melting of the spin order parameter $|\bm J|$. Dashed lines indicate exponential fits of the evolution $J(t)=J_{\infty}+\Delta J e^{-t/\tau}$. (b) The change of double occupancy upon photoexcitation. An inverse temperature $\beta=50$ is chosen for the initial equilibrium state.}
\label{tri_decay}
\end{figure}

\begin{figure}
\includegraphics[scale=0.7]{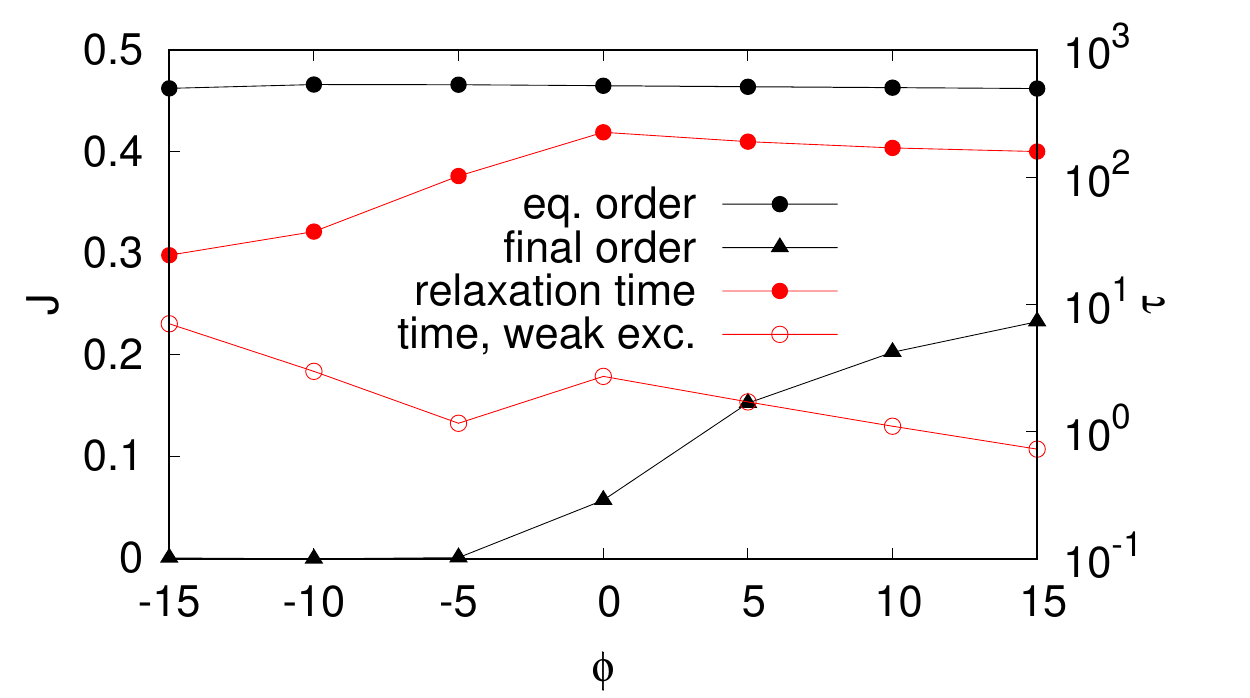}
\caption{The equilibrium spin order and relaxation time for different SOC angle $\phi$'s in the triangular lattice. The relaxation time (red dots) and final order ($|\bm J(\infty)|$, triangles) are obtained from the fits shown in Fig.~\ref{tri_decay}(a). The hollow red points indicate the relaxation time for a weaker excitation $A_0=0.3$ for comparison. }
\label{order}
\end{figure}

\subsection{Demagnetization of the $120^\circ$ AFM order}
We start with the $120^\circ$ AFM state and apply the same electric pulse as in Sect.~\ref{sq}, leading to the spin dynamics in Fig.~\ref{tri_decay} for $\phi=-15^\circ,-10^{\circ},\ldots, 15^\circ$. The $120^\circ$ relation between neighboring spins is assumed during the dynamics. Fig.~\ref{tri_decay}(b) shows the evolution of double occupancy $d=\langle n_{\uparrow}n_{\downarrow}\rangle$. Notice that the initially created double occupancy 
$\Delta d$ is always around $\Delta d \sim 1\times10^{-2}$ 
but, different from the previous situation where changing $\phi$ leaves dynamics intact,  the subsequent time-evolution of the order $|\bm J|$ deviates for different SOC angle $\phi$'s, as shown in Fig.~\ref{tri_decay}(a). In general, a positive SOC angle $\phi>0$ suppresses melting, whereas a negative $\phi<0$ enhances melting. In particular, the excitation density, measured by the initial increase of double occupancy, is almost identical for all values of $\phi$ in Fig.~\ref{tri_decay}(b), while the spin dynamics still changes dramatically from $\phi=0$ to $15^\circ$. Although the equilibrium order is essentially unaffected for these values of $\phi$, see Fig.~\ref{order}, the relaxation time $\tau$ drops for both positive and negative $\phi$'s. Indeed, compared with Fig.~\ref{tri_decay}(a), one can see the trend of the slow down close to the threshold of a complete melting (around $\phi=0^\circ$, as shown in Fig.~\ref{order}). 
%ME
A dependence of the spin dynamics on the angle $\phi$ is observed for weaker excitations, see the curve for $A_0=0.3$ in Fig.~\ref{order} (the order parameter is only weakly perturbed and reduced by up to about $10\%$ for this excitation density). In this case the relaxation occurs in a very short period, and stronger melting is approached for negative $\phi$'s.
%The similar behavior is observed for weaker excitations, where the $A_0=0.3$ and the $|\bm J_{\rm eff}|$ is reduced up to about $10\%$. In this case the relaxation occurs in a very short period, and a strong melting is approached for negative $\phi$'s.

\subsection{Controlling demagnetization by modulating spin-charge coupling}

In the case of Sr$_2$IrO$_4$, the spin dynamics with SOC can be mapped to dynamics without SOC, using a static rotation of the local spin basis responsible for the spin canting. The melting of spin order is controlled only through tuning the bandwidth by changing $\theta$. In the triangular lattice, however, the SOC strongly affects the dynamics, although it does not induce any further spin canting on top of the original $120^\circ$ order. To understand this phenomenology, we come back to the spin-charge coupling mechanism discussed in Sec.~\ref{sq} and examine the effect of SOC on this mechanism.

\begin{figure}
\includegraphics[scale=0.7]{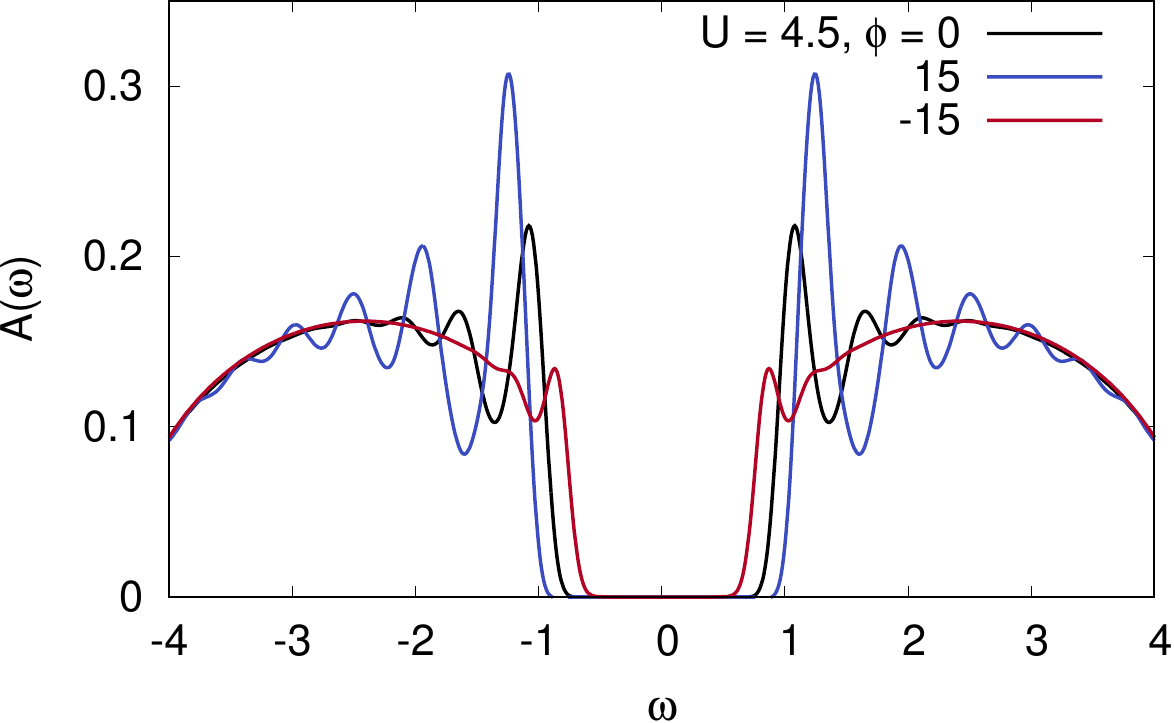}
\caption{The spin-averaged spectral function $A(\omega)$ with ($\phi=\pm15^\circ$) and without SOC ($\phi=0$) in the triangular lattice. The oscillatory peaks correspond to the spin polaron excitations. }
\label{spec}
\end{figure}

In equilibrium, the coupling between spin and charge is known to create spin polarons in AFM Mott insulators \cite{Brinkman1970,Dagotto1994,Brunner2000,DalConte2015,Grusdt2018}. These are doublons/holons hybridizing with magnons through their fluctuating motions. The polarons reveal themselves by the localization peaks \cite{strack1992, sangiovanni2006} in the spectral function $A(\omega)=-\operatorname{Im}G^R(\omega)/\pi$. We obtain $A(\omega)$ by calculating the Fourier transform of the equilibrium retarded Green's function $G^R(t-t')$, see Fig.~\ref{spec}. We consider $\phi=0$ and $\pm15^\circ$ to emphasize the difference. While the equilibrium spin orders are all saturated with $|\bm J|\sim 0.45$ (see the black curve in Fig.~\ref{order}), the spectral function is very different: for $\phi=15^\circ$, the system has a slightly larger Mott gap and significantly stronger spin polaron peaks than $\phi=0$, implying enhanced localization of the doublons and holons. $\phi=-15^\circ$, however, leads to an opposite trend, with smaller Mott gap and suppressed spin polaron peaks. Note that the chirality of the $120^\circ$ order is fixed in our simulation, and the positive and negative $\phi$'s results in different dynamics. The larger gap at $\phi>0$ suppresses the photoinduced creation of the charge excitations, leading to slightly lower double occupancy in Fig.~\ref{tri_decay}(b). The stronger localization of charge excitations further 
%ME
reduces 
%reduce 
the efficiency of the charge excitations to destruct the AFM order, suppressing the demagnetization (see Fig.~\ref{tri_decay}(a)). The opposite trend for $\phi<0$ is explained with the same argument. 

To further understand the behavior of the spin-polaron physics, we evaluate the exchange energy $I_{\rm ex}\langle \tilde{\bm J}_{1}\cdot \tilde{ \bm J}_{2}\rangle$ and show that it is generally modified by the SOC even if the spin order parameter is unaffected. Indeed, for the triangular lattice, assuming the counterclockwise $120^\circ$ order and saturated effective spin $|\bm J_{1}|=\hbar/2$, the exchange energy is simply given by $I_{\rm ex}\hbar^2\cos(2\pi/3 - \phi)/4$ 
%ME
within a mean-field approximation, implying enhanced (suppressed) exchange energy for $\phi>0$ ($\phi<0$). 
%under mean-field approximations. 
This explains the features of Fig.~\ref{spec}. Indeed, since the motion of charge excitations creates spin excitations (mismatched neighboring spins), whose energy cost is proportional to the exchange energy, the enhanced exchange energy under $\phi>0$ then makes charge excitations more difficult to move around, leading to a weaker melting of the spin order. On the other hand, $\phi<0$ reduces the energy cost and leads to enhanced demagnetization. 

%ME
Within the DMFT simulation, the chirality of the $120^\circ$ order is fixed by hand. In reality, the stable configuration is determined by its  free energy, while the other chirality may exist as a metastable domain. The above estimate of the mean field energy suggests that the energetically favorable chirality is the one in which the localization of the electrons is increased. As a consequence, one can say that the SOC in the triangular geometry enhances both the spin-polaron effects and weakens the effect of photo-doping on the spin order in the thermodynamically most stable state. If both chiralities would exist in a lattice, photo-doping could be used to selectively affect one of them. 

\section{Conclusion}
\label{conc}
In this article, we studied the ultrafast spin dynamics of the Mott insulators with spin-orbit coupling and demonstrated the possibility of tuning ultrafast demagnetization with changing SOC in two prototypical examples, including the one-band Hubbard model on a square lattice and on a triangular lattice. The models are generally relevant to transition metal compounds and cold atom systems \cite{struck2011,struck2014,hamner2015}. We find that the light-induced melting of effective spin moments in Sr$_2$IrO$_4$ strongly relies on its crystal structure parameter, i.e. the rotation angle $\theta$ of the IrO$_6$ octahedra, which determines the SOC strength and controls the bandwidth. The SOC-induced canting angle is nevertheless intact during the demagnetization. In the triangular lattice, the SOC strongly affects the melting of the $120^\circ$ order. Specifically, an SOC angle $\phi>0$ suppresses the melting, whereas $\phi<0$ enhances the melting, allowing for flexible tunability of the dynamics. This trend can be explained by the modulation of spin exchange energy by the SOC, which increases (decreases) the energy cost for a charge excitation to hop in the AFM background when the SOC angle is consistent (inconsistent) with the chirality of the spin order. This effect can be observed in terms of the spin polaron peaks in the spectrum. 

The square and triangular geometries can be identified as representative of two more general scenarios how the SOC influences the spin dynamics. The first type gives rise to unfrustrated spin canting, such as in Sr$_2$IrO$_4$, and the Hamiltonian can be mapped to a non-SOC Hubbard model. The dynamics of the canted AFM is then related to that of the collinear AFM in the non-SOC case with a time-independent spin canting, leading to a fixed canting angle. The second type, considered in the triangular lattice, results in frustrated spin canting. In this case, the spin exchange energy is effectively modified by the SOC, modulating the coupling between the AFM order and charge excitations and controlling the spin dynamics. 

We have studied the SOC physics in minimal models with a single band. The underpinning mechanisms can be readily generalized to more complex situations, in which the SOC can result in canting of the combined spin and orbital orders and modify the exchange coupling in the compass model \cite{jackeli2009}. In the future, further studies can be carried out for the spin-orbit multi-band Hubbard models with, for example, $t_{2g}$ orbital degeneracy and Hund's coupling \cite{khomskii2014}. The SOC engineering of the order-parameter dynamics can be generalized to intertwined spin, orbital, and charge orders and may have crucial consequences on photoinduced hidden phases \cite{li2018nat}. For this prospect, the effects of nonlocal fluctuations and electron-lattice coupling can be studied with newly developed methods, such as $GW$+DMFT \cite{sun2002, biermann2003} in the nonequilibrium regime \cite{golez2019} and an exact treatment of the phonon degree of freedom \cite{grandi2020}. These methods can be combined with the steady-state theory of photodoping \cite{li2020} to assess the long-time behavior of the order parameters after the excitation.

\begin{acknowledgments}
We thank D.~Afanasiev for useful discussions. We were supported by ERC starting Grant No. 716648. The authors gratefully acknowledge the compute resources and support provided by the Erlangen Regional Computing Center (RRZE). 
\end{acknowledgments}
\bibliography{soc.bib}

\begin{thebibliography}{62}
\expandafter\ifx\csname natexlab\endcsname\relax\def\natexlab#1{#1}\fi
\expandafter\ifx\csname bibnamefont\endcsname\relax
  \def\bibnamefont#1{#1}\fi
\expandafter\ifx\csname bibfnamefont\endcsname\relax
  \def\bibfnamefont#1{#1}\fi
\expandafter\ifx\csname citenamefont\endcsname\relax
  \def\citenamefont#1{#1}\fi
\expandafter\ifx\csname url\endcsname\relax
  \def\url#1{\texttt{#1}}\fi
\expandafter\ifx\csname urlprefix\endcsname\relax\def\urlprefix{URL }\fi
\providecommand{\bibinfo}[2]{#2}
\providecommand{\eprint}[2][]{\url{#2}}

\bibitem[{\citenamefont{Basov et~al.}(2017)\citenamefont{Basov, Averitt, and
  Hsieh}}]{basov2017}
\bibinfo{author}{\bibfnamefont{D.}~\bibnamefont{Basov}},
  \bibinfo{author}{\bibfnamefont{R.}~\bibnamefont{Averitt}}, \bibnamefont{and}
  \bibinfo{author}{\bibfnamefont{D.}~\bibnamefont{Hsieh}},
  \bibinfo{journal}{Nat. Mater.} \textbf{\bibinfo{volume}{16}},
  \bibinfo{pages}{1077} (\bibinfo{year}{2017}).

\bibitem[{\citenamefont{Orenstein}(2012)}]{orenstein2012}
\bibinfo{author}{\bibfnamefont{J.}~\bibnamefont{Orenstein}},
  \bibinfo{journal}{Physics Today} \textbf{\bibinfo{volume}{65}},
  \bibinfo{pages}{44} (\bibinfo{year}{2012}).

\bibitem[{\citenamefont{Kirilyuk et~al.}(2010)\citenamefont{Kirilyuk, Kimel,
  and Rasing}}]{kirilyuk2010}
\bibinfo{author}{\bibfnamefont{A.}~\bibnamefont{Kirilyuk}},
  \bibinfo{author}{\bibfnamefont{A.~V.} \bibnamefont{Kimel}}, \bibnamefont{and}
  \bibinfo{author}{\bibfnamefont{T.}~\bibnamefont{Rasing}},
  \bibinfo{journal}{Rev. Mod. Phys.} \textbf{\bibinfo{volume}{82}},
  \bibinfo{pages}{2731} (\bibinfo{year}{2010}).

\bibitem[{\citenamefont{Beaurepaire et~al.}(1996)\citenamefont{Beaurepaire,
  Merle, Daunois, and Bigot}}]{beaurepaire1996}
\bibinfo{author}{\bibfnamefont{E.}~\bibnamefont{Beaurepaire}},
  \bibinfo{author}{\bibfnamefont{J.-C.} \bibnamefont{Merle}},
  \bibinfo{author}{\bibfnamefont{A.}~\bibnamefont{Daunois}}, \bibnamefont{and}
  \bibinfo{author}{\bibfnamefont{J.-Y.} \bibnamefont{Bigot}},
  \bibinfo{journal}{Phys. Rev. Lett.} \textbf{\bibinfo{volume}{76}},
  \bibinfo{pages}{4250} (\bibinfo{year}{1996}),
  \urlprefix\url{https://link.aps.org/doi/10.1103/PhysRevLett.76.4250}.

\bibitem[{\citenamefont{Kimel et~al.}(2002)\citenamefont{Kimel, Pisarev,
  Hohlfeld, and Rasing}}]{kimel2002}
\bibinfo{author}{\bibfnamefont{A.~V.} \bibnamefont{Kimel}},
  \bibinfo{author}{\bibfnamefont{R.~V.} \bibnamefont{Pisarev}},
  \bibinfo{author}{\bibfnamefont{J.}~\bibnamefont{Hohlfeld}}, \bibnamefont{and}
  \bibinfo{author}{\bibfnamefont{T.}~\bibnamefont{Rasing}},
  \bibinfo{journal}{Phys. Rev. Lett.} \textbf{\bibinfo{volume}{89}},
  \bibinfo{pages}{287401} (\bibinfo{year}{2002}),
  \urlprefix\url{https://link.aps.org/doi/10.1103/PhysRevLett.89.287401}.

\bibitem[{\citenamefont{Imada et~al.}(1998)\citenamefont{Imada, Fujimori, and
  Tokura}}]{imada1998}
\bibinfo{author}{\bibfnamefont{M.}~\bibnamefont{Imada}},
  \bibinfo{author}{\bibfnamefont{A.}~\bibnamefont{Fujimori}}, \bibnamefont{and}
  \bibinfo{author}{\bibfnamefont{Y.}~\bibnamefont{Tokura}},
  \bibinfo{journal}{Rev. Mod. Phys.} \textbf{\bibinfo{volume}{70}},
  \bibinfo{pages}{1039} (\bibinfo{year}{1998}),
  \urlprefix\url{https://link.aps.org/doi/10.1103/RevModPhys.70.1039}.

\bibitem[{\citenamefont{Zhang et~al.}(2019)\citenamefont{Zhang, Staar,
  Kozhevnikov, Wang, Trinastic, Schulthess, and Cheng}}]{Zhang2019}
\bibinfo{author}{\bibfnamefont{L.}~\bibnamefont{Zhang}},
  \bibinfo{author}{\bibfnamefont{P.}~\bibnamefont{Staar}},
  \bibinfo{author}{\bibfnamefont{A.}~\bibnamefont{Kozhevnikov}},
  \bibinfo{author}{\bibfnamefont{Y.-P.} \bibnamefont{Wang}},
  \bibinfo{author}{\bibfnamefont{J.}~\bibnamefont{Trinastic}},
  \bibinfo{author}{\bibfnamefont{T.}~\bibnamefont{Schulthess}},
  \bibnamefont{and} \bibinfo{author}{\bibfnamefont{H.-P.} \bibnamefont{Cheng}},
  \bibinfo{journal}{Phys. Rev. B} \textbf{\bibinfo{volume}{100}},
  \bibinfo{pages}{035104} (\bibinfo{year}{2019}),
  \urlprefix\url{https://link.aps.org/doi/10.1103/PhysRevB.100.035104}.

\bibitem[{\citenamefont{Held et~al.}(2001)\citenamefont{Held, Keller, Eyert,
  Vollhardt, and Anisimov}}]{Held2001}
\bibinfo{author}{\bibfnamefont{K.}~\bibnamefont{Held}},
  \bibinfo{author}{\bibfnamefont{G.}~\bibnamefont{Keller}},
  \bibinfo{author}{\bibfnamefont{V.}~\bibnamefont{Eyert}},
  \bibinfo{author}{\bibfnamefont{D.}~\bibnamefont{Vollhardt}},
  \bibnamefont{and} \bibinfo{author}{\bibfnamefont{V.~I.}
  \bibnamefont{Anisimov}}, \bibinfo{journal}{Phys. Rev. Lett.}
  \textbf{\bibinfo{volume}{86}}, \bibinfo{pages}{5345} (\bibinfo{year}{2001}),
  \urlprefix\url{https://link.aps.org/doi/10.1103/PhysRevLett.86.5345}.

\bibitem[{\citenamefont{Cai et~al.}(2016)\citenamefont{Cai, Ruan, Peng, Ye, Li,
  Hao, Zhou, Lee, and Wang}}]{Cai2016}
\bibinfo{author}{\bibfnamefont{P.}~\bibnamefont{Cai}},
  \bibinfo{author}{\bibfnamefont{W.}~\bibnamefont{Ruan}},
  \bibinfo{author}{\bibfnamefont{Y.}~\bibnamefont{Peng}},
  \bibinfo{author}{\bibfnamefont{C.}~\bibnamefont{Ye}},
  \bibinfo{author}{\bibfnamefont{X.}~\bibnamefont{Li}},
  \bibinfo{author}{\bibfnamefont{Z.}~\bibnamefont{Hao}},
  \bibinfo{author}{\bibfnamefont{X.}~\bibnamefont{Zhou}},
  \bibinfo{author}{\bibfnamefont{D.-H.} \bibnamefont{Lee}}, \bibnamefont{and}
  \bibinfo{author}{\bibfnamefont{Y.}~\bibnamefont{Wang}},
  \bibinfo{journal}{Nature Physics} \textbf{\bibinfo{volume}{12}},
  \bibinfo{pages}{1047} (\bibinfo{year}{2016}), ISSN \bibinfo{issn}{1745-2481},
  \urlprefix\url{http://dx.doi.org/10.1038/nphys3840}.

\bibitem[{\citenamefont{Mor et~al.}(2017)\citenamefont{Mor, Herzog, Golez,
  Werner, Eckstein, Katayama, Nohara, Takagi, Mizokawa, Monney
  et~al.}}]{mor2017}
\bibinfo{author}{\bibfnamefont{S.}~\bibnamefont{Mor}},
  \bibinfo{author}{\bibfnamefont{M.}~\bibnamefont{Herzog}},
  \bibinfo{author}{\bibfnamefont{D.}~\bibnamefont{Golez}},
  \bibinfo{author}{\bibfnamefont{P.}~\bibnamefont{Werner}},
  \bibinfo{author}{\bibfnamefont{M.}~\bibnamefont{Eckstein}},
  \bibinfo{author}{\bibfnamefont{N.}~\bibnamefont{Katayama}},
  \bibinfo{author}{\bibfnamefont{M.}~\bibnamefont{Nohara}},
  \bibinfo{author}{\bibfnamefont{H.}~\bibnamefont{Takagi}},
  \bibinfo{author}{\bibfnamefont{T.}~\bibnamefont{Mizokawa}},
  \bibinfo{author}{\bibfnamefont{C.}~\bibnamefont{Monney}},
  \bibnamefont{et~al.}, \bibinfo{journal}{Phys. Rev. Lett.}
  \textbf{\bibinfo{volume}{119}}, \bibinfo{pages}{086401}
  (\bibinfo{year}{2017}),
  \urlprefix\url{https://link.aps.org/doi/10.1103/PhysRevLett.119.086401}.

\bibitem[{\citenamefont{Beaud et~al.}(2014)\citenamefont{Beaud, Caviezel,
  Mariager, Rettig, Ingold, Dornes, Huang, Johnson, Radovic, Huber
  et~al.}}]{beaud2014}
\bibinfo{author}{\bibfnamefont{P.}~\bibnamefont{Beaud}},
  \bibinfo{author}{\bibfnamefont{A.}~\bibnamefont{Caviezel}},
  \bibinfo{author}{\bibfnamefont{S.}~\bibnamefont{Mariager}},
  \bibinfo{author}{\bibfnamefont{L.}~\bibnamefont{Rettig}},
  \bibinfo{author}{\bibfnamefont{G.}~\bibnamefont{Ingold}},
  \bibinfo{author}{\bibfnamefont{C.}~\bibnamefont{Dornes}},
  \bibinfo{author}{\bibfnamefont{S.}~\bibnamefont{Huang}},
  \bibinfo{author}{\bibfnamefont{J.}~\bibnamefont{Johnson}},
  \bibinfo{author}{\bibfnamefont{M.}~\bibnamefont{Radovic}},
  \bibinfo{author}{\bibfnamefont{T.}~\bibnamefont{Huber}},
  \bibnamefont{et~al.}, \bibinfo{journal}{Nature materials}
  \textbf{\bibinfo{volume}{13}}, \bibinfo{pages}{923} (\bibinfo{year}{2014}).

\bibitem[{\citenamefont{Lenar\ifmmode \check{c}\else
  \v{c}\fi{}i\ifmmode~\check{c}\else \v{c}\fi{} and
  Prelov\ifmmode~\check{s}\else \v{s}\fi{}ek}(2013)}]{lenarcic2013}
\bibinfo{author}{\bibfnamefont{Z.}~\bibnamefont{Lenar\ifmmode \check{c}\else
  \v{c}\fi{}i\ifmmode~\check{c}\else \v{c}\fi{}}} \bibnamefont{and}
  \bibinfo{author}{\bibfnamefont{P.}~\bibnamefont{Prelov\ifmmode~\check{s}\else
  \v{s}\fi{}ek}}, \bibinfo{journal}{Phys. Rev. Lett.}
  \textbf{\bibinfo{volume}{111}}, \bibinfo{pages}{016401}
  (\bibinfo{year}{2013}),
  \urlprefix\url{https://link.aps.org/doi/10.1103/PhysRevLett.111.016401}.

\bibitem[{\citenamefont{Golez et~al.}(2014)\citenamefont{Golez, Bonca,
  Mierzejewski, and Vidmar}}]{Golez2014}
\bibinfo{author}{\bibfnamefont{D.}~\bibnamefont{Golez}},
  \bibinfo{author}{\bibfnamefont{J.}~\bibnamefont{Bonca}},
  \bibinfo{author}{\bibfnamefont{M.}~\bibnamefont{Mierzejewski}},
  \bibnamefont{and} \bibinfo{author}{\bibfnamefont{L.}~\bibnamefont{Vidmar}},
  \bibinfo{journal}{Phys. Rev. B} \textbf{\bibinfo{volume}{89}},
  \bibinfo{pages}{165118} (\bibinfo{year}{2014}),
  \urlprefix\url{https://link.aps.org/doi/10.1103/PhysRevB.89.165118}.

\bibitem[{\citenamefont{Balzer et~al.}(2015)\citenamefont{Balzer, Wolf,
  McCulloch, Werner, and Eckstein}}]{balzer2015}
\bibinfo{author}{\bibfnamefont{K.}~\bibnamefont{Balzer}},
  \bibinfo{author}{\bibfnamefont{F.~A.} \bibnamefont{Wolf}},
  \bibinfo{author}{\bibfnamefont{I.~P.} \bibnamefont{McCulloch}},
  \bibinfo{author}{\bibfnamefont{P.}~\bibnamefont{Werner}}, \bibnamefont{and}
  \bibinfo{author}{\bibfnamefont{M.}~\bibnamefont{Eckstein}},
  \bibinfo{journal}{Phys. Rev. X} \textbf{\bibinfo{volume}{5}},
  \bibinfo{pages}{031039} (\bibinfo{year}{2015}),
  \urlprefix\url{https://link.aps.org/doi/10.1103/PhysRevX.5.031039}.

\bibitem[{\citenamefont{Dal~Conte et~al.}(2015)\citenamefont{Dal~Conte, Vidmar,
  Golez, Mierzejewski, Soavi, Peli, Banfi, Ferrini, Comin, Ludbrook
  et~al.}}]{DalConte2015}
\bibinfo{author}{\bibfnamefont{S.}~\bibnamefont{Dal~Conte}},
  \bibinfo{author}{\bibfnamefont{L.}~\bibnamefont{Vidmar}},
  \bibinfo{author}{\bibfnamefont{D.}~\bibnamefont{Golez}},
  \bibinfo{author}{\bibfnamefont{M.}~\bibnamefont{Mierzejewski}},
  \bibinfo{author}{\bibfnamefont{G.}~\bibnamefont{Soavi}},
  \bibinfo{author}{\bibfnamefont{S.}~\bibnamefont{Peli}},
  \bibinfo{author}{\bibfnamefont{F.}~\bibnamefont{Banfi}},
  \bibinfo{author}{\bibfnamefont{G.}~\bibnamefont{Ferrini}},
  \bibinfo{author}{\bibfnamefont{R.}~\bibnamefont{Comin}},
  \bibinfo{author}{\bibfnamefont{B.~M.} \bibnamefont{Ludbrook}},
  \bibnamefont{et~al.}, \bibinfo{journal}{Nature Physics}
  \textbf{\bibinfo{volume}{11}}, \bibinfo{pages}{421} (\bibinfo{year}{2015}),
  \urlprefix\url{https://doi.org/10.1038/nphys3265}.

\bibitem[{\citenamefont{Aoki et~al.}(2014)\citenamefont{Aoki, Tsuji, Eckstein,
  Kollar, Oka, and Werner}}]{aoki2014}
\bibinfo{author}{\bibfnamefont{H.}~\bibnamefont{Aoki}},
  \bibinfo{author}{\bibfnamefont{N.}~\bibnamefont{Tsuji}},
  \bibinfo{author}{\bibfnamefont{M.}~\bibnamefont{Eckstein}},
  \bibinfo{author}{\bibfnamefont{M.}~\bibnamefont{Kollar}},
  \bibinfo{author}{\bibfnamefont{T.}~\bibnamefont{Oka}}, \bibnamefont{and}
  \bibinfo{author}{\bibfnamefont{P.}~\bibnamefont{Werner}},
  \bibinfo{journal}{Rev. Mod. Phys.} \textbf{\bibinfo{volume}{86}},
  \bibinfo{pages}{779} (\bibinfo{year}{2014}),
  \urlprefix\url{https://link.aps.org/doi/10.1103/RevModPhys.86.779}.

\bibitem[{\citenamefont{Georges et~al.}(1996)\citenamefont{Georges, Kotliar,
  Krauth, and Rozenberg}}]{georges1996}
\bibinfo{author}{\bibfnamefont{A.}~\bibnamefont{Georges}},
  \bibinfo{author}{\bibfnamefont{G.}~\bibnamefont{Kotliar}},
  \bibinfo{author}{\bibfnamefont{W.}~\bibnamefont{Krauth}}, \bibnamefont{and}
  \bibinfo{author}{\bibfnamefont{M.~J.} \bibnamefont{Rozenberg}},
  \bibinfo{journal}{Rev. Mod. Phys.} \textbf{\bibinfo{volume}{68}},
  \bibinfo{pages}{13} (\bibinfo{year}{1996}),
  \urlprefix\url{https://link.aps.org/doi/10.1103/RevModPhys.68.13}.

\bibitem[{\citenamefont{Brinkman and Rice}(1970)}]{Brinkman1970}
\bibinfo{author}{\bibfnamefont{W.~F.} \bibnamefont{Brinkman}} \bibnamefont{and}
  \bibinfo{author}{\bibfnamefont{T.~M.} \bibnamefont{Rice}},
  \bibinfo{journal}{Phys. Rev. B} \textbf{\bibinfo{volume}{2}},
  \bibinfo{pages}{1324} (\bibinfo{year}{1970}),
  \urlprefix\url{https://link.aps.org/doi/10.1103/PhysRevB.2.1324}.

\bibitem[{\citenamefont{Dagotto}(1994)}]{Dagotto1994}
\bibinfo{author}{\bibfnamefont{E.}~\bibnamefont{Dagotto}},
  \bibinfo{journal}{Rev. Mod. Phys.} \textbf{\bibinfo{volume}{66}},
  \bibinfo{pages}{763} (\bibinfo{year}{1994}),
  \urlprefix\url{https://link.aps.org/doi/10.1103/RevModPhys.66.763}.

\bibitem[{\citenamefont{Brunner et~al.}(2000)\citenamefont{Brunner, Assaad, and
  Muramatsu}}]{Brunner2000}
\bibinfo{author}{\bibfnamefont{M.}~\bibnamefont{Brunner}},
  \bibinfo{author}{\bibfnamefont{F.~F.} \bibnamefont{Assaad}},
  \bibnamefont{and}
  \bibinfo{author}{\bibfnamefont{A.}~\bibnamefont{Muramatsu}},
  \bibinfo{journal}{Phys. Rev. B} \textbf{\bibinfo{volume}{62}},
  \bibinfo{pages}{15480} (\bibinfo{year}{2000}),
  \urlprefix\url{https://link.aps.org/doi/10.1103/PhysRevB.62.15480}.

\bibitem[{\citenamefont{Sangiovanni et~al.}(2006)\citenamefont{Sangiovanni,
  Toschi, Koch, Held, Capone, Castellani, Gunnarsson, Mo, Allen, Kim
  et~al.}}]{sangiovanni2006}
\bibinfo{author}{\bibfnamefont{G.}~\bibnamefont{Sangiovanni}},
  \bibinfo{author}{\bibfnamefont{A.}~\bibnamefont{Toschi}},
  \bibinfo{author}{\bibfnamefont{E.}~\bibnamefont{Koch}},
  \bibinfo{author}{\bibfnamefont{K.}~\bibnamefont{Held}},
  \bibinfo{author}{\bibfnamefont{M.}~\bibnamefont{Capone}},
  \bibinfo{author}{\bibfnamefont{C.}~\bibnamefont{Castellani}},
  \bibinfo{author}{\bibfnamefont{O.}~\bibnamefont{Gunnarsson}},
  \bibinfo{author}{\bibfnamefont{S.-K.} \bibnamefont{Mo}},
  \bibinfo{author}{\bibfnamefont{J.~W.} \bibnamefont{Allen}},
  \bibinfo{author}{\bibfnamefont{H.-D.} \bibnamefont{Kim}},
  \bibnamefont{et~al.}, \bibinfo{journal}{Phys. Rev. B}
  \textbf{\bibinfo{volume}{73}}, \bibinfo{pages}{205121}
  (\bibinfo{year}{2006}),
  \urlprefix\url{https://link.aps.org/doi/10.1103/PhysRevB.73.205121}.

\bibitem[{\citenamefont{Grusdt et~al.}(2018)\citenamefont{Grusdt,
  K\'anasz-Nagy, Bohrdt, Chiu, Ji, Greiner, Greif, and Demler}}]{Grusdt2018}
\bibinfo{author}{\bibfnamefont{F.}~\bibnamefont{Grusdt}},
  \bibinfo{author}{\bibfnamefont{M.}~\bibnamefont{K\'anasz-Nagy}},
  \bibinfo{author}{\bibfnamefont{A.}~\bibnamefont{Bohrdt}},
  \bibinfo{author}{\bibfnamefont{C.~S.} \bibnamefont{Chiu}},
  \bibinfo{author}{\bibfnamefont{G.}~\bibnamefont{Ji}},
  \bibinfo{author}{\bibfnamefont{M.}~\bibnamefont{Greiner}},
  \bibinfo{author}{\bibfnamefont{D.}~\bibnamefont{Greif}}, \bibnamefont{and}
  \bibinfo{author}{\bibfnamefont{E.}~\bibnamefont{Demler}},
  \bibinfo{journal}{Phys. Rev. X} \textbf{\bibinfo{volume}{8}},
  \bibinfo{pages}{011046} (\bibinfo{year}{2018}),
  \urlprefix\url{https://link.aps.org/doi/10.1103/PhysRevX.8.011046}.

\bibitem[{\citenamefont{Srivastava and Singh}(2005)}]{Srivastava2005}
\bibinfo{author}{\bibfnamefont{P.}~\bibnamefont{Srivastava}} \bibnamefont{and}
  \bibinfo{author}{\bibfnamefont{A.}~\bibnamefont{Singh}},
  \bibinfo{journal}{Phys. Rev. B} \textbf{\bibinfo{volume}{72}},
  \bibinfo{pages}{224409} (\bibinfo{year}{2005}),
  \urlprefix\url{https://link.aps.org/doi/10.1103/PhysRevB.72.224409}.

\bibitem[{\citenamefont{Tohyama}(2006)}]{Tohyama2006}
\bibinfo{author}{\bibfnamefont{T.}~\bibnamefont{Tohyama}},
  \bibinfo{journal}{Phys. Rev. B} \textbf{\bibinfo{volume}{74}},
  \bibinfo{pages}{113108} (\bibinfo{year}{2006}),
  \urlprefix\url{https://link.aps.org/doi/10.1103/PhysRevB.74.113108}.

\bibitem[{\citenamefont{Hamad et~al.}(2006)\citenamefont{Hamad, Manuel,
  Martinez, and Trumper}}]{Hamad2006}
\bibinfo{author}{\bibfnamefont{I.~J.} \bibnamefont{Hamad}},
  \bibinfo{author}{\bibfnamefont{L.~O.} \bibnamefont{Manuel}},
  \bibinfo{author}{\bibfnamefont{G.}~\bibnamefont{Martinez}}, \bibnamefont{and}
  \bibinfo{author}{\bibfnamefont{A.~E.} \bibnamefont{Trumper}},
  \bibinfo{journal}{Phys. Rev. B} \textbf{\bibinfo{volume}{74}},
  \bibinfo{pages}{094417} (\bibinfo{year}{2006}),
  \urlprefix\url{https://link.aps.org/doi/10.1103/PhysRevB.74.094417}.

\bibitem[{\citenamefont{Hamad et~al.}(2008)\citenamefont{Hamad, Trumper,
  Feiguin, and Manuel}}]{Hamad2008}
\bibinfo{author}{\bibfnamefont{I.~J.} \bibnamefont{Hamad}},
  \bibinfo{author}{\bibfnamefont{A.~E.} \bibnamefont{Trumper}},
  \bibinfo{author}{\bibfnamefont{A.~E.} \bibnamefont{Feiguin}},
  \bibnamefont{and} \bibinfo{author}{\bibfnamefont{L.~O.}
  \bibnamefont{Manuel}}, \bibinfo{journal}{Phys. Rev. B}
  \textbf{\bibinfo{volume}{77}}, \bibinfo{pages}{014410}
  (\bibinfo{year}{2008}),
  \urlprefix\url{https://link.aps.org/doi/10.1103/PhysRevB.77.014410}.

\bibitem[{\citenamefont{L\"auchli and Poilblanc}(2004)}]{Laeuchli2004}
\bibinfo{author}{\bibfnamefont{A.}~\bibnamefont{L\"auchli}} \bibnamefont{and}
  \bibinfo{author}{\bibfnamefont{D.}~\bibnamefont{Poilblanc}},
  \bibinfo{journal}{Phys. Rev. Lett.} \textbf{\bibinfo{volume}{92}},
  \bibinfo{pages}{236404} (\bibinfo{year}{2004}),
  \urlprefix\url{https://link.aps.org/doi/10.1103/PhysRevLett.92.236404}.

\bibitem[{\citenamefont{Shibata et~al.}(1999)\citenamefont{Shibata, Tohyama,
  and Maekawa}}]{Shibata1999}
\bibinfo{author}{\bibfnamefont{Y.}~\bibnamefont{Shibata}},
  \bibinfo{author}{\bibfnamefont{T.}~\bibnamefont{Tohyama}}, \bibnamefont{and}
  \bibinfo{author}{\bibfnamefont{S.}~\bibnamefont{Maekawa}},
  \bibinfo{journal}{Phys. Rev. B} \textbf{\bibinfo{volume}{59}},
  \bibinfo{pages}{1840} (\bibinfo{year}{1999}),
  \urlprefix\url{https://link.aps.org/doi/10.1103/PhysRevB.59.1840}.

\bibitem[{\citenamefont{Capriotti et~al.}(1999)\citenamefont{Capriotti,
  Trumper, and Sorella}}]{Capriotti1999}
\bibinfo{author}{\bibfnamefont{L.}~\bibnamefont{Capriotti}},
  \bibinfo{author}{\bibfnamefont{A.~E.} \bibnamefont{Trumper}},
  \bibnamefont{and} \bibinfo{author}{\bibfnamefont{S.}~\bibnamefont{Sorella}},
  \bibinfo{journal}{Phys. Rev. Lett.} \textbf{\bibinfo{volume}{82}},
  \bibinfo{pages}{3899} (\bibinfo{year}{1999}),
  \urlprefix\url{https://link.aps.org/doi/10.1103/PhysRevLett.82.3899}.

\bibitem[{\citenamefont{White and Chernyshev}(2007)}]{White2007}
\bibinfo{author}{\bibfnamefont{S.~R.} \bibnamefont{White}} \bibnamefont{and}
  \bibinfo{author}{\bibfnamefont{A.~L.} \bibnamefont{Chernyshev}},
  \bibinfo{journal}{Phys. Rev. Lett.} \textbf{\bibinfo{volume}{99}},
  \bibinfo{pages}{127004} (\bibinfo{year}{2007}),
  \urlprefix\url{https://link.aps.org/doi/10.1103/PhysRevLett.99.127004}.

\bibitem[{\citenamefont{Iida et~al.}(2019)\citenamefont{Iida, Yoshida, Okabe,
  Katayama, Ishii, Koda, Inamura, Murai, Ishikado, Kadono et~al.}}]{Iida2019}
\bibinfo{author}{\bibfnamefont{K.}~\bibnamefont{Iida}},
  \bibinfo{author}{\bibfnamefont{H.}~\bibnamefont{Yoshida}},
  \bibinfo{author}{\bibfnamefont{H.}~\bibnamefont{Okabe}},
  \bibinfo{author}{\bibfnamefont{N.}~\bibnamefont{Katayama}},
  \bibinfo{author}{\bibfnamefont{Y.}~\bibnamefont{Ishii}},
  \bibinfo{author}{\bibfnamefont{A.}~\bibnamefont{Koda}},
  \bibinfo{author}{\bibfnamefont{Y.}~\bibnamefont{Inamura}},
  \bibinfo{author}{\bibfnamefont{N.}~\bibnamefont{Murai}},
  \bibinfo{author}{\bibfnamefont{M.}~\bibnamefont{Ishikado}},
  \bibinfo{author}{\bibfnamefont{R.}~\bibnamefont{Kadono}},
  \bibnamefont{et~al.}, \bibinfo{journal}{Scientific Reports}
  \textbf{\bibinfo{volume}{9}}, \bibinfo{pages}{1826} (\bibinfo{year}{2019}),
  \urlprefix\url{https://doi.org/10.1038/s41598-018-36123-7}.

\bibitem[{\citenamefont{Bittner et~al.}(2020)\citenamefont{Bittner, Golez,
  Eckstein, and Werner}}]{Bittner2020}
\bibinfo{author}{\bibfnamefont{N.}~\bibnamefont{Bittner}},
  \bibinfo{author}{\bibfnamefont{D.}~\bibnamefont{Golez}},
  \bibinfo{author}{\bibfnamefont{M.}~\bibnamefont{Eckstein}}, \bibnamefont{and}
  \bibinfo{author}{\bibfnamefont{P.}~\bibnamefont{Werner}},
  \emph{\bibinfo{title}{Effects of frustration on the nonequilibrium dynamics
  of photo-excited lattice systems}} (\bibinfo{year}{2020}),
  \eprint{2005.11722}.

\bibitem[{\citenamefont{Afanasiev et~al.}(2019)\citenamefont{Afanasiev,
  Gatilova, Groenendijk, Ivanov, Gibert, Gariglio, Mentink, Li, Dasari,
  Eckstein et~al.}}]{afanasiev2019}
\bibinfo{author}{\bibfnamefont{D.}~\bibnamefont{Afanasiev}},
  \bibinfo{author}{\bibfnamefont{A.}~\bibnamefont{Gatilova}},
  \bibinfo{author}{\bibfnamefont{D.~J.} \bibnamefont{Groenendijk}},
  \bibinfo{author}{\bibfnamefont{B.~A.} \bibnamefont{Ivanov}},
  \bibinfo{author}{\bibfnamefont{M.}~\bibnamefont{Gibert}},
  \bibinfo{author}{\bibfnamefont{S.}~\bibnamefont{Gariglio}},
  \bibinfo{author}{\bibfnamefont{J.}~\bibnamefont{Mentink}},
  \bibinfo{author}{\bibfnamefont{J.}~\bibnamefont{Li}},
  \bibinfo{author}{\bibfnamefont{N.}~\bibnamefont{Dasari}},
  \bibinfo{author}{\bibfnamefont{M.}~\bibnamefont{Eckstein}},
  \bibnamefont{et~al.}, \bibinfo{journal}{Phys. Rev. X}
  \textbf{\bibinfo{volume}{9}}, \bibinfo{pages}{021020} (\bibinfo{year}{2019}),
  \urlprefix\url{https://link.aps.org/doi/10.1103/PhysRevX.9.021020}.

\bibitem[{\citenamefont{Jackeli and Khaliullin}(2009)}]{jackeli2009}
\bibinfo{author}{\bibfnamefont{G.}~\bibnamefont{Jackeli}} \bibnamefont{and}
  \bibinfo{author}{\bibfnamefont{G.}~\bibnamefont{Khaliullin}},
  \bibinfo{journal}{Phys. Rev. Lett.} \textbf{\bibinfo{volume}{102}},
  \bibinfo{pages}{017205} (\bibinfo{year}{2009}),
  \urlprefix\url{https://link.aps.org/doi/10.1103/PhysRevLett.102.017205}.

\bibitem[{\citenamefont{Jin et~al.}(2009)\citenamefont{Jin, Jeong, Ozaki, and
  Yu}}]{jin2009}
\bibinfo{author}{\bibfnamefont{H.}~\bibnamefont{Jin}},
  \bibinfo{author}{\bibfnamefont{H.}~\bibnamefont{Jeong}},
  \bibinfo{author}{\bibfnamefont{T.}~\bibnamefont{Ozaki}}, \bibnamefont{and}
  \bibinfo{author}{\bibfnamefont{J.}~\bibnamefont{Yu}}, \bibinfo{journal}{Phys.
  Rev. B} \textbf{\bibinfo{volume}{80}}, \bibinfo{pages}{075112}
  (\bibinfo{year}{2009}),
  \urlprefix\url{https://link.aps.org/doi/10.1103/PhysRevB.80.075112}.

\bibitem[{\citenamefont{Wang and Senthil}(2011)}]{wang2011}
\bibinfo{author}{\bibfnamefont{F.}~\bibnamefont{Wang}} \bibnamefont{and}
  \bibinfo{author}{\bibfnamefont{T.}~\bibnamefont{Senthil}},
  \bibinfo{journal}{Phys. Rev. Lett.} \textbf{\bibinfo{volume}{106}},
  \bibinfo{pages}{136402} (\bibinfo{year}{2011}),
  \urlprefix\url{https://link.aps.org/doi/10.1103/PhysRevLett.106.136402}.

\bibitem[{\citenamefont{Kim et~al.}(2012)\citenamefont{Kim, Khaliullin, and
  Min}}]{kim2012}
\bibinfo{author}{\bibfnamefont{B.~H.} \bibnamefont{Kim}},
  \bibinfo{author}{\bibfnamefont{G.}~\bibnamefont{Khaliullin}},
  \bibnamefont{and} \bibinfo{author}{\bibfnamefont{B.~I.} \bibnamefont{Min}},
  \bibinfo{journal}{Phys. Rev. Lett.} \textbf{\bibinfo{volume}{109}},
  \bibinfo{pages}{167205} (\bibinfo{year}{2012}),
  \urlprefix\url{https://link.aps.org/doi/10.1103/PhysRevLett.109.167205}.

\bibitem[{\citenamefont{Yan et~al.}(2015)\citenamefont{Yan, Ren, Xu, Xie, Tao,
  Choi, Lee, Choi, Zhang, and Feng}}]{yan2015}
\bibinfo{author}{\bibfnamefont{Y.~J.} \bibnamefont{Yan}},
  \bibinfo{author}{\bibfnamefont{M.~Q.} \bibnamefont{Ren}},
  \bibinfo{author}{\bibfnamefont{H.~C.} \bibnamefont{Xu}},
  \bibinfo{author}{\bibfnamefont{B.~P.} \bibnamefont{Xie}},
  \bibinfo{author}{\bibfnamefont{R.}~\bibnamefont{Tao}},
  \bibinfo{author}{\bibfnamefont{H.~Y.} \bibnamefont{Choi}},
  \bibinfo{author}{\bibfnamefont{N.}~\bibnamefont{Lee}},
  \bibinfo{author}{\bibfnamefont{Y.~J.} \bibnamefont{Choi}},
  \bibinfo{author}{\bibfnamefont{T.}~\bibnamefont{Zhang}}, \bibnamefont{and}
  \bibinfo{author}{\bibfnamefont{D.~L.} \bibnamefont{Feng}},
  \bibinfo{journal}{Phys. Rev. X} \textbf{\bibinfo{volume}{5}},
  \bibinfo{pages}{041018} (\bibinfo{year}{2015}),
  \urlprefix\url{https://link.aps.org/doi/10.1103/PhysRevX.5.041018}.

\bibitem[{\citenamefont{Dean et~al.}(2016)\citenamefont{Dean, Cao, Liu, Wall,
  Zhu, Mankowsky, Thampy, Chen, Vale, Casa et~al.}}]{dean2016}
\bibinfo{author}{\bibfnamefont{M.~P.} \bibnamefont{Dean}},
  \bibinfo{author}{\bibfnamefont{Y.}~\bibnamefont{Cao}},
  \bibinfo{author}{\bibfnamefont{X.}~\bibnamefont{Liu}},
  \bibinfo{author}{\bibfnamefont{S.}~\bibnamefont{Wall}},
  \bibinfo{author}{\bibfnamefont{D.}~\bibnamefont{Zhu}},
  \bibinfo{author}{\bibfnamefont{R.}~\bibnamefont{Mankowsky}},
  \bibinfo{author}{\bibfnamefont{V.}~\bibnamefont{Thampy}},
  \bibinfo{author}{\bibfnamefont{X.}~\bibnamefont{Chen}},
  \bibinfo{author}{\bibfnamefont{J.~G.} \bibnamefont{Vale}},
  \bibinfo{author}{\bibfnamefont{D.}~\bibnamefont{Casa}}, \bibnamefont{et~al.},
  \bibinfo{journal}{Nature materials} \textbf{\bibinfo{volume}{15}},
  \bibinfo{pages}{601} (\bibinfo{year}{2016}).

\bibitem[{\citenamefont{Versteeg et~al.}(2020)\citenamefont{Versteeg,
  Chiocchetta, Sekiguchi, Aldea, Sahasrabudhe, Budzinauskas, Wang, Tsurkan,
  Loidl, Khomskii et~al.}}]{versteeg2020}
\bibinfo{author}{\bibfnamefont{R.}~\bibnamefont{Versteeg}},
  \bibinfo{author}{\bibfnamefont{A.}~\bibnamefont{Chiocchetta}},
  \bibinfo{author}{\bibfnamefont{F.}~\bibnamefont{Sekiguchi}},
  \bibinfo{author}{\bibfnamefont{A.}~\bibnamefont{Aldea}},
  \bibinfo{author}{\bibfnamefont{A.}~\bibnamefont{Sahasrabudhe}},
  \bibinfo{author}{\bibfnamefont{K.}~\bibnamefont{Budzinauskas}},
  \bibinfo{author}{\bibfnamefont{Z.}~\bibnamefont{Wang}},
  \bibinfo{author}{\bibfnamefont{V.}~\bibnamefont{Tsurkan}},
  \bibinfo{author}{\bibfnamefont{A.}~\bibnamefont{Loidl}},
  \bibinfo{author}{\bibfnamefont{D.}~\bibnamefont{Khomskii}},
  \bibnamefont{et~al.}, \bibinfo{journal}{arXiv preprint arXiv:2005.14189}
  (\bibinfo{year}{2020}).

\bibitem[{\citenamefont{Hamner et~al.}(2015)\citenamefont{Hamner, Zhang,
  Khamehchi, Davis, and Engels}}]{hamner2015}
\bibinfo{author}{\bibfnamefont{C.}~\bibnamefont{Hamner}},
  \bibinfo{author}{\bibfnamefont{Y.}~\bibnamefont{Zhang}},
  \bibinfo{author}{\bibfnamefont{M.}~\bibnamefont{Khamehchi}},
  \bibinfo{author}{\bibfnamefont{M.~J.} \bibnamefont{Davis}}, \bibnamefont{and}
  \bibinfo{author}{\bibfnamefont{P.}~\bibnamefont{Engels}},
  \bibinfo{journal}{Phys. Rev. Lett.} \textbf{\bibinfo{volume}{114}},
  \bibinfo{pages}{070401} (\bibinfo{year}{2015}).

\bibitem[{\citenamefont{Bonesteel et~al.}(1992)\citenamefont{Bonesteel, Rice,
  and Zhang}}]{bonesteel1992}
\bibinfo{author}{\bibfnamefont{N.~E.} \bibnamefont{Bonesteel}},
  \bibinfo{author}{\bibfnamefont{T.~M.} \bibnamefont{Rice}}, \bibnamefont{and}
  \bibinfo{author}{\bibfnamefont{F.~C.} \bibnamefont{Zhang}},
  \bibinfo{journal}{Phys. Rev. Lett.} \textbf{\bibinfo{volume}{68}},
  \bibinfo{pages}{2684} (\bibinfo{year}{1992}),
  \urlprefix\url{https://link.aps.org/doi/10.1103/PhysRevLett.68.2684}.

\bibitem[{\citenamefont{Shitade et~al.}(2009)\citenamefont{Shitade, Katsura,
  Kune\ifmmode~\check{s}\else \v{s}\fi{}, Qi, Zhang, and
  Nagaosa}}]{shitade2009}
\bibinfo{author}{\bibfnamefont{A.}~\bibnamefont{Shitade}},
  \bibinfo{author}{\bibfnamefont{H.}~\bibnamefont{Katsura}},
  \bibinfo{author}{\bibfnamefont{J.}~\bibnamefont{Kune\ifmmode~\check{s}\else
  \v{s}\fi{}}}, \bibinfo{author}{\bibfnamefont{X.-L.} \bibnamefont{Qi}},
  \bibinfo{author}{\bibfnamefont{S.-C.} \bibnamefont{Zhang}}, \bibnamefont{and}
  \bibinfo{author}{\bibfnamefont{N.}~\bibnamefont{Nagaosa}},
  \bibinfo{journal}{Phys. Rev. Lett.} \textbf{\bibinfo{volume}{102}},
  \bibinfo{pages}{256403} (\bibinfo{year}{2009}),
  \urlprefix\url{https://link.aps.org/doi/10.1103/PhysRevLett.102.256403}.

\bibitem[{\citenamefont{Sch{\"u}ler et~al.}(2020)\citenamefont{Sch{\"u}ler,
  Golez, Murakami, Bittner, Herrmann, Strand, Werner, and
  Eckstein}}]{schuler2020}
\bibinfo{author}{\bibfnamefont{M.}~\bibnamefont{Sch{\"u}ler}},
  \bibinfo{author}{\bibfnamefont{D.}~\bibnamefont{Golez}},
  \bibinfo{author}{\bibfnamefont{Y.}~\bibnamefont{Murakami}},
  \bibinfo{author}{\bibfnamefont{N.}~\bibnamefont{Bittner}},
  \bibinfo{author}{\bibfnamefont{A.}~\bibnamefont{Herrmann}},
  \bibinfo{author}{\bibfnamefont{H.~U.} \bibnamefont{Strand}},
  \bibinfo{author}{\bibfnamefont{P.}~\bibnamefont{Werner}}, \bibnamefont{and}
  \bibinfo{author}{\bibfnamefont{M.}~\bibnamefont{Eckstein}},
  \bibinfo{journal}{Computer Physics Communications}
  \textbf{\bibinfo{volume}{257}}, \bibinfo{pages}{107484}
  (\bibinfo{year}{2020}).

\bibitem[{\citenamefont{Kawamura}(1998)}]{kawamura1998}
\bibinfo{author}{\bibfnamefont{H.}~\bibnamefont{Kawamura}},
  \bibinfo{journal}{Journal of Physics: Condensed Matter}
  \textbf{\bibinfo{volume}{10}}, \bibinfo{pages}{4707} (\bibinfo{year}{1998}).

\bibitem[{\citenamefont{Moriya}(1960)}]{moriya1960}
\bibinfo{author}{\bibfnamefont{T.}~\bibnamefont{Moriya}},
  \bibinfo{journal}{Phys. Rev.} \textbf{\bibinfo{volume}{120}},
  \bibinfo{pages}{91} (\bibinfo{year}{1960}),
  \urlprefix\url{https://link.aps.org/doi/10.1103/PhysRev.120.91}.

\bibitem[{\citenamefont{Liu et~al.}(2015)\citenamefont{Liu, Khmelevskyi, Kim,
  Marsman, Li, Chen, Sarma, Kresse, and Franchini}}]{liu2015}
\bibinfo{author}{\bibfnamefont{P.}~\bibnamefont{Liu}},
  \bibinfo{author}{\bibfnamefont{S.}~\bibnamefont{Khmelevskyi}},
  \bibinfo{author}{\bibfnamefont{B.}~\bibnamefont{Kim}},
  \bibinfo{author}{\bibfnamefont{M.}~\bibnamefont{Marsman}},
  \bibinfo{author}{\bibfnamefont{D.}~\bibnamefont{Li}},
  \bibinfo{author}{\bibfnamefont{X.-Q.} \bibnamefont{Chen}},
  \bibinfo{author}{\bibfnamefont{D.~D.} \bibnamefont{Sarma}},
  \bibinfo{author}{\bibfnamefont{G.}~\bibnamefont{Kresse}}, \bibnamefont{and}
  \bibinfo{author}{\bibfnamefont{C.}~\bibnamefont{Franchini}},
  \bibinfo{journal}{Phys. Rev. B} \textbf{\bibinfo{volume}{92}},
  \bibinfo{pages}{054428} (\bibinfo{year}{2015}),
  \urlprefix\url{https://link.aps.org/doi/10.1103/PhysRevB.92.054428}.

\bibitem[{\citenamefont{Iwai et~al.}(2003)\citenamefont{Iwai, Ono, Maeda,
  Matsuzaki, Kishida, Okamoto, and Tokura}}]{iwai2003}
\bibinfo{author}{\bibfnamefont{S.}~\bibnamefont{Iwai}},
  \bibinfo{author}{\bibfnamefont{M.}~\bibnamefont{Ono}},
  \bibinfo{author}{\bibfnamefont{A.}~\bibnamefont{Maeda}},
  \bibinfo{author}{\bibfnamefont{H.}~\bibnamefont{Matsuzaki}},
  \bibinfo{author}{\bibfnamefont{H.}~\bibnamefont{Kishida}},
  \bibinfo{author}{\bibfnamefont{H.}~\bibnamefont{Okamoto}}, \bibnamefont{and}
  \bibinfo{author}{\bibfnamefont{Y.}~\bibnamefont{Tokura}},
  \bibinfo{journal}{Phys. Rev. Lett.} \textbf{\bibinfo{volume}{91}},
  \bibinfo{pages}{057401} (\bibinfo{year}{2003}),
  \urlprefix\url{https://link.aps.org/doi/10.1103/PhysRevLett.91.057401}.

\bibitem[{\citenamefont{Okamoto et~al.}(2010)\citenamefont{Okamoto, Miyagoe,
  Kobayashi, Uemura, Nishioka, Matsuzaki, Sawa, and Tokura}}]{okamoto2010}
\bibinfo{author}{\bibfnamefont{H.}~\bibnamefont{Okamoto}},
  \bibinfo{author}{\bibfnamefont{T.}~\bibnamefont{Miyagoe}},
  \bibinfo{author}{\bibfnamefont{K.}~\bibnamefont{Kobayashi}},
  \bibinfo{author}{\bibfnamefont{H.}~\bibnamefont{Uemura}},
  \bibinfo{author}{\bibfnamefont{H.}~\bibnamefont{Nishioka}},
  \bibinfo{author}{\bibfnamefont{H.}~\bibnamefont{Matsuzaki}},
  \bibinfo{author}{\bibfnamefont{A.}~\bibnamefont{Sawa}}, \bibnamefont{and}
  \bibinfo{author}{\bibfnamefont{Y.}~\bibnamefont{Tokura}},
  \bibinfo{journal}{Phys. Rev. B} \textbf{\bibinfo{volume}{82}},
  \bibinfo{pages}{060513} (\bibinfo{year}{2010}),
  \urlprefix\url{https://link.aps.org/doi/10.1103/PhysRevB.82.060513}.

\bibitem[{\citenamefont{Sensarma et~al.}(2010)\citenamefont{Sensarma, Pekker,
  Altman, Demler, Strohmaier, Greif, J\"ordens, Tarruell, Moritz, and
  Esslinger}}]{sensarma2010}
\bibinfo{author}{\bibfnamefont{R.}~\bibnamefont{Sensarma}},
  \bibinfo{author}{\bibfnamefont{D.}~\bibnamefont{Pekker}},
  \bibinfo{author}{\bibfnamefont{E.}~\bibnamefont{Altman}},
  \bibinfo{author}{\bibfnamefont{E.}~\bibnamefont{Demler}},
  \bibinfo{author}{\bibfnamefont{N.}~\bibnamefont{Strohmaier}},
  \bibinfo{author}{\bibfnamefont{D.}~\bibnamefont{Greif}},
  \bibinfo{author}{\bibfnamefont{R.}~\bibnamefont{J\"ordens}},
  \bibinfo{author}{\bibfnamefont{L.}~\bibnamefont{Tarruell}},
  \bibinfo{author}{\bibfnamefont{H.}~\bibnamefont{Moritz}}, \bibnamefont{and}
  \bibinfo{author}{\bibfnamefont{T.}~\bibnamefont{Esslinger}},
  \bibinfo{journal}{Phys. Rev. B} \textbf{\bibinfo{volume}{82}},
  \bibinfo{pages}{224302} (\bibinfo{year}{2010}),
  \urlprefix\url{https://link.aps.org/doi/10.1103/PhysRevB.82.224302}.

\bibitem[{\citenamefont{Eckstein and Werner}(2014)}]{eckstein2014}
\bibinfo{author}{\bibfnamefont{M.}~\bibnamefont{Eckstein}} \bibnamefont{and}
  \bibinfo{author}{\bibfnamefont{P.}~\bibnamefont{Werner}},
  \bibinfo{journal}{Phys. Rev. Lett.} \textbf{\bibinfo{volume}{113}},
  \bibinfo{pages}{076405} (\bibinfo{year}{2014}),
  \urlprefix\url{https://link.aps.org/doi/10.1103/PhysRevLett.113.076405}.

\bibitem[{\citenamefont{Mitrano et~al.}(2014)\citenamefont{Mitrano, Cotugno,
  Clark, Singla, Kaiser, St\"ahler, Beyer, Dressel, Baldassarre, Nicoletti
  et~al.}}]{mitrano2014}
\bibinfo{author}{\bibfnamefont{M.}~\bibnamefont{Mitrano}},
  \bibinfo{author}{\bibfnamefont{G.}~\bibnamefont{Cotugno}},
  \bibinfo{author}{\bibfnamefont{S.~R.} \bibnamefont{Clark}},
  \bibinfo{author}{\bibfnamefont{R.}~\bibnamefont{Singla}},
  \bibinfo{author}{\bibfnamefont{S.}~\bibnamefont{Kaiser}},
  \bibinfo{author}{\bibfnamefont{J.}~\bibnamefont{St\"ahler}},
  \bibinfo{author}{\bibfnamefont{R.}~\bibnamefont{Beyer}},
  \bibinfo{author}{\bibfnamefont{M.}~\bibnamefont{Dressel}},
  \bibinfo{author}{\bibfnamefont{L.}~\bibnamefont{Baldassarre}},
  \bibinfo{author}{\bibfnamefont{D.}~\bibnamefont{Nicoletti}},
  \bibnamefont{et~al.}, \bibinfo{journal}{Phys. Rev. Lett.}
  \textbf{\bibinfo{volume}{112}}, \bibinfo{pages}{117801}
  (\bibinfo{year}{2014}),
  \urlprefix\url{https://link.aps.org/doi/10.1103/PhysRevLett.112.117801}.

\bibitem[{\citenamefont{Strack and Vollhardt}(1992)}]{strack1992}
\bibinfo{author}{\bibfnamefont{R.}~\bibnamefont{Strack}} \bibnamefont{and}
  \bibinfo{author}{\bibfnamefont{D.}~\bibnamefont{Vollhardt}},
  \bibinfo{journal}{Phys. Rev. B} \textbf{\bibinfo{volume}{46}},
  \bibinfo{pages}{13852} (\bibinfo{year}{1992}),
  \urlprefix\url{https://link.aps.org/doi/10.1103/PhysRevB.46.13852}.

\bibitem[{\citenamefont{Struck et~al.}(2011)\citenamefont{Struck,
  {\"O}lschl{\"a}ger, Le~Targat, Soltan-Panahi, Eckardt, Lewenstein,
  Windpassinger, and Sengstock}}]{struck2011}
\bibinfo{author}{\bibfnamefont{J.}~\bibnamefont{Struck}},
  \bibinfo{author}{\bibfnamefont{C.}~\bibnamefont{{\"O}lschl{\"a}ger}},
  \bibinfo{author}{\bibfnamefont{R.}~\bibnamefont{Le~Targat}},
  \bibinfo{author}{\bibfnamefont{P.}~\bibnamefont{Soltan-Panahi}},
  \bibinfo{author}{\bibfnamefont{A.}~\bibnamefont{Eckardt}},
  \bibinfo{author}{\bibfnamefont{M.}~\bibnamefont{Lewenstein}},
  \bibinfo{author}{\bibfnamefont{P.}~\bibnamefont{Windpassinger}},
  \bibnamefont{and}
  \bibinfo{author}{\bibfnamefont{K.}~\bibnamefont{Sengstock}},
  \bibinfo{journal}{Science} \textbf{\bibinfo{volume}{333}},
  \bibinfo{pages}{996} (\bibinfo{year}{2011}).

\bibitem[{\citenamefont{Struck et~al.}(2014)\citenamefont{Struck, Simonet, and
  Sengstock}}]{struck2014}
\bibinfo{author}{\bibfnamefont{J.}~\bibnamefont{Struck}},
  \bibinfo{author}{\bibfnamefont{J.}~\bibnamefont{Simonet}}, \bibnamefont{and}
  \bibinfo{author}{\bibfnamefont{K.}~\bibnamefont{Sengstock}},
  \bibinfo{journal}{Phys. Rev. A} \textbf{\bibinfo{volume}{90}},
  \bibinfo{pages}{031601} (\bibinfo{year}{2014}),
  \urlprefix\url{https://link.aps.org/doi/10.1103/PhysRevA.90.031601}.

\bibitem[{\citenamefont{Khomskii}(2014)}]{khomskii2014}
\bibinfo{author}{\bibfnamefont{D.}~\bibnamefont{Khomskii}},
  \emph{\bibinfo{title}{Transition metal compounds}}
  (\bibinfo{publisher}{Cambridge University Press}, \bibinfo{year}{2014}).

\bibitem[{\citenamefont{Li et~al.}(2018)\citenamefont{Li, Strand, Werner, and
  Eckstein}}]{li2018nat}
\bibinfo{author}{\bibfnamefont{J.}~\bibnamefont{Li}},
  \bibinfo{author}{\bibfnamefont{H.~U.} \bibnamefont{Strand}},
  \bibinfo{author}{\bibfnamefont{P.}~\bibnamefont{Werner}}, \bibnamefont{and}
  \bibinfo{author}{\bibfnamefont{M.}~\bibnamefont{Eckstein}},
  \bibinfo{journal}{Nat. Commun.} \textbf{\bibinfo{volume}{9}},
  \bibinfo{pages}{1} (\bibinfo{year}{2018}).

\bibitem[{\citenamefont{Sun and Kotliar}(2002)}]{sun2002}
\bibinfo{author}{\bibfnamefont{P.}~\bibnamefont{Sun}} \bibnamefont{and}
  \bibinfo{author}{\bibfnamefont{G.}~\bibnamefont{Kotliar}},
  \bibinfo{journal}{Phys. Rev. B} \textbf{\bibinfo{volume}{66}},
  \bibinfo{pages}{085120} (\bibinfo{year}{2002}),
  \urlprefix\url{https://link.aps.org/doi/10.1103/PhysRevB.66.085120}.

\bibitem[{\citenamefont{Biermann et~al.}(2003)\citenamefont{Biermann,
  Aryasetiawan, and Georges}}]{biermann2003}
\bibinfo{author}{\bibfnamefont{S.}~\bibnamefont{Biermann}},
  \bibinfo{author}{\bibfnamefont{F.}~\bibnamefont{Aryasetiawan}},
  \bibnamefont{and} \bibinfo{author}{\bibfnamefont{A.}~\bibnamefont{Georges}},
  \bibinfo{journal}{Phys. Rev. Lett.} \textbf{\bibinfo{volume}{90}},
  \bibinfo{pages}{086402} (\bibinfo{year}{2003}),
  \urlprefix\url{https://link.aps.org/doi/10.1103/PhysRevLett.90.086402}.

\bibitem[{\citenamefont{Golez et~al.}(2019)\citenamefont{Golez, Eckstein, and
  Werner}}]{golez2019}
\bibinfo{author}{\bibfnamefont{D.}~\bibnamefont{Golez}},
  \bibinfo{author}{\bibfnamefont{M.}~\bibnamefont{Eckstein}}, \bibnamefont{and}
  \bibinfo{author}{\bibfnamefont{P.}~\bibnamefont{Werner}},
  \bibinfo{journal}{Phys. Rev. B} \textbf{\bibinfo{volume}{100}},
  \bibinfo{pages}{235117} (\bibinfo{year}{2019}),
  \urlprefix\url{https://link.aps.org/doi/10.1103/PhysRevB.100.235117}.

\bibitem[{\citenamefont{Grandi et~al.}(2020)\citenamefont{Grandi, Li, and
  Eckstein}}]{grandi2020}
\bibinfo{author}{\bibfnamefont{F.}~\bibnamefont{Grandi}},
  \bibinfo{author}{\bibfnamefont{J.}~\bibnamefont{Li}}, \bibnamefont{and}
  \bibinfo{author}{\bibfnamefont{M.}~\bibnamefont{Eckstein}},
  \emph{\bibinfo{title}{Ultrafast mott transition driven by nonlinear phonons}}
  (\bibinfo{year}{2020}), \eprint{arXiv 2005.14100}.

\bibitem[{\citenamefont{Li and Eckstein}(2020)}]{li2020}
\bibinfo{author}{\bibfnamefont{J.}~\bibnamefont{Li}} \bibnamefont{and}
  \bibinfo{author}{\bibfnamefont{M.}~\bibnamefont{Eckstein}},
  \bibinfo{journal}{arXiv preprint arXiv:2007.12511}  (\bibinfo{year}{2020}).

\end{thebibliography}
\end{document}